\documentclass[12pt,aaspp4]{aastex}

\newcommand\kms{\ifmmode{\rm km\th s^{-1}}\else km\th s$^{-1}$\fi}
\newcommand\msun{\ifmmode{M_{\odot}}\else $M_{\odot}$\fi}
\newcommand\rsun{\ifmmode{R_{\odot}}\else $R_{\odot}$\fi}

\begin{document}

\title{INFRARED PHOTOMETRY OF NGC 6791\footnotemark[1]}

\author{Bruce W.\ Carney\altaffilmark{2}, Jae-Woo Lee\altaffilmark{3}, 
\and Benjamin Dodson\altaffilmark{2}}

\footnotetext[1]{Based on observations made with the Mayall 4-meter
telescope of the National Optical Astronomy Observatory}
\altaffiltext{2}{Department of Physics \& Astronomy,
University of North Carolina, Chapel Hill, NC 27599-3255;
bruce@physics.unc.edu; benjadod@email.unc.edu}
\altaffiltext{3}{Department of Astronomy \& Space Science,
Astrophysical Research Center for the Structure and
Evolution of the Cosmos,
Sejong University, 98 Gunja-Dong, Gwangjin-Gu, Seoul, 143-747;
jaewoo@arcsec.sejong.ac.kr}

\begin{abstract}
We present deep $JHK$ photometry of the old and metal-rich
open cluster NGC~6791. The photometry reaches below the main
sequence turn-off to $K \approx 16.5$ mag. We combine our
photometry with that from Stetson, Bruntt, \& Grundahl (2003)
to provide color-magnitude diagrams showing $K$ vs.\ $J-K$,
$K$ vs.\ $V-K$, and $V$ vs.\ $V-K$. We study the slope of
the red giant branch in the infrared, but find that it is not
a useful metallicity indicator for the cluster, nor any metal-rich
cluster that lacks a well-populated red giant branch, because
it is not linear, as has often been assumed, in $K$ vs.\ $J-K$.
The mean color of the red horizontal branch/red clump
stars provide an estimate the cluster reddening, E($B-V$) = $0.14 \pm 0.04$
mag for [Fe/H] = $+0.4 \pm 0.1$. The
mean magnitudes of these stars also provide a good distance
estimate, ($m-M$)$_{0}$ = $13.07 \pm 0.04$. Finally, we find
that the isochrones of Yi, Kim, \& Demarque (2003)
provide optimal fits in $V$ vs.\ $B-V$ and $V-K$ and $K$ vs. $J-K$
and $V-K$ for such values if [Fe/H] lies between +0.3 and +0.5
(with a slight preference for +0.5) and ages between 9 Gyrs
([Fe/H] = +0.3) and 7.5 Gyrs ([Fe/H] = +0.5).

\end{abstract}

\keywords{Galaxy: disk --- open clusters: individual (NGC~6791) ---
stars: abundances --- infrared: stars}

\section{INTRODUCTION}

NGC~6791 is a remarkable star cluster. It is one of the very few
known open clusters whose age approaches that of globular clusters
and the halo generally, with recent age estimates ranging from 8~Gyrs
(Chaboyer, Green, \& Liebert 1999) to 10~Gyrs (Tripicco et al.\ 1995)
to 12~Gyrs (Stetson, Bruntt, \& Grundahl 2003; hereafter SBG2003).
The cluster is therefore a rare fossil find, and is one of the
few samples where we can study stars that formed very early in the
evolution of the Galaxy's disk. Despite its age, its heavy elements
appear to be more abundant than in the much younger Sun and
comparable age stars and clusters at the solar Galactocentric distance.

The abundances are difficult to measure in this moderately distant
and moderately extinguished cluster, and metallicity estimates or
measures have been provided using three general methods, all of them
summarized well by SBG2003. Metallicity
``indicators" are so named because they
must be calibrated using other clusters or
stars whose abundances have been \underline{measured} using high-resolution
spectroscopy. Canterna et al.\ (1986), using Washington system
photometry, and
Friel \& Janes (1993), using $DDO$ photometry, pointed the way
to a ``super-solar metallicity" for NGC~6791.
Low- and moderate-resolution spectroscopy,
likewise calibrated using stars whose abundances have been determined
using high-resolution spectroscopy, have also implied high metallicities
(Friel et al.\ 20002; Worthey \& Jowett 2003; see also
Cole et al.\ 2004). The model isochrones employed
to derive estimates of the cluster's age may also help estimate
the cluster's chemical composition, although the problem is then
multi-parametric. Chaboyer et al.\ (1999) and SBG2003 have both
used this approach to infer heavy element mass fractions of at
least twice solar. The most direct method to determine the
cluster's chemical composition is, of course, high-resolution
spectrscopy. This is not as simple as it sounds. The cluster
is relatively distant, with ($m-M$)$_{0}$ estimates ranging from
12.6 to 13.6 (!). Thus even its most luminous stars are relatively
faint. However, such stars are also quite cool, and the combination
of low effective temperature and high metallicity conspire to make
such spectra difficult to evaluate. This is (partly) what led
Peterson \& Green (1998) to study star 2-17, which is apparently
a cluster member, despite being a ``blue horizontal branch" star;
that is, a fairly hot star with a rather thin hydrogen envelope
and a hydrogen-burning shell above a helium-burning core.
The existence of such stars in NCG~6791 adds spice to an already
sumptuous cluster, and also show that the metallicity is indeed
very high, [Fe/H] $\approx\ +0.4 \pm 0.1$. The cluster's reddening
estimates are also somewhat uncertain, but a careful study of field
stars in the direction of the cluster revealed E($B-V$) = $0.10 \pm 0.02$
(Montgomery, Janes, \& Phelps 1994).
Taylor (2001) has critically reviewed the
metallicity estimates and concluded that [Fe/H] most probably
lies in the range of +0.16 to +0.44.

So NGC~6791 truly is remarkable: it somehow formed long before the Sun,
and not too long after the onset of star formation had begun in the
Galactic disk, yet its star-forming environment had already produced twice
as many heavy elements per unit mass as had the environment out
of which the Sun and similar stars formed several billion years later.
For this reason alone, NGC~6791 deserves extensive study.

There is a second reason for an observational focus on NGC~6791. Because
it is both old and metal-rich, it can play a critical role in
calibrating the diverse ``secondary" metallicity indicators, such
as low- to moderate-resolution spectroscopy or photometry.

We argue that infrared photometry of the cluster can adds valuable
information. The slope of the red giant branch in
$K$ vs.\ $J-K$ is reddening- and distant-independent, yet has
been shown to be a metallicity indicator [Kuchinski et al.\ (1995);
Kuchinski \& Frogel (1995); Ferraro et al.\ (2000); Ivanov \& Borissova (2002);
Valenti, Ferraro, and Origlia (2004a; hereafter VFO); Valenti et al.\ 2004b].
Metal-rich globular clusters also appear to have de-reddened mean colors of
their red horizontal branch stars that are insensitive to metallicity,
and claims have also been made that
their mean $K$-band luminosities do not depend strongly
on metallicity (Kuchinski et al.\ 1995; Kuchinski \& Frogel 1995; but see
Grocholski \& Sarajedini 2002). Thus such stars in NGC~6791 may
provide independent reddening and distance estimates. Infrared
colors such as ($J-K$) are less affected by line
blanketing and hence show reduced sensitivity to metallicity than optical
colors. (See Alonso, Arribas, \& Martinez-Roger 1996 for a discussion
of color-temperature relations for main sequence stars, and
Alonso, Arribas, \& Martinez 1999 for a similar discussion of
giant stars).
Finally, the
combination of optical and infrared photometry provides very long
wavelength baselines and colors that have special sensitivity to
temperature independent of metallicity, such as $V-K$.

\section{OBSERVATIONS \& DATA REDUCTIONS}

NGC~6791 was observed on the night of June 15-16, 1997 using the Kitt Peak
National Observatory 4-meter Mayall telescope and the IRIM camera. IRIM
contained a 256 $\times$ 256 pixels HgCdTe NICMOS3 detector,
with an image scale of
roughly 0.6\arcsec\ pixel$^{-1}$, and a field of view of 154\arcsec\ $\times$
154\arcsec.
The night was clear throughout, with the full width at half maxmimum
image quality ranged from 0.8\arcsec\ to 1.0\arcsec.

The HgCdTe detector was operated with a gain setting of 10.46 electrons
per ADU, and the full well capacity was equivalent to about 400,000 electrons.
The readnoise was 35 electrons, and the typical dark level was 2 electrons
(not discernable at our gain setting). The device was non-linear
for the brightest objects, the effect size reaching about 6\% at
350,000 electrons. We calibrated the non-linearity pixel-by-pixel
by taking a lengthy
series of dome flats. Exposures of 1, 3, 1, 10, 1, 15, 1, 20, 1, 30, and 1
seconds were taken through the $J$ filter. The interspersed 1-sec frames
enabled us to monitor any changes in the calibration lamp illumination.
Dark frames were also taken every day at all the exposure times employed for the
observing run. (See Lee \& Carney 2002 for additional discussion of
the corrections for non-linearity, especially their Figure~1.)

The cluster was observed using 0.5, 1.0, and 10-second integration
times, with each
recorded image resulting from the co-addition of two such observations.
For each pointing of the telescope, a five-position raster pattern was
employed to dither the images on 5\arcsec\ scales. This was done primarily to
produce a smoother sky background, which dominates the photometric precision
for most of our program stars. Two such rasters were done at each
pointing, yielding a set of images for each field with a cumulative
exposure time of 200 seconds.
After each cluster field had been observed,
a single set of 5-position rasters was made of a field in a region of
lower stellar density to improve the sky background measurement.

Five cluster fields were observed. One was centered on the cluster's
central position. Each of the remaining four fields
were offset by roughly 75\arcsec\ in both
right ascension and declination in the northeast, northwest, southeast,
and southwest directions. Thus all four
offset fields overlapped different quadrants of the cluster central field,
enabling us to better register the images to a common grid,
and to provide somewhat improved observations for the central
field since this observation pattern effectively doubled the integration
time for all of the central stars.

Seven standard stars were observed during the night, with one of them
followed from airmass 1.2 to 2.0 to determine the atmospheric extinction.
The standard stars
were taken from the list of faint $JHK$ standard stars for UKIRT
(Casali \& Hawarden 1992), and
ranged from 9.76 to 13.12 in $K$, and had a good
range in color, $J-K$ = $-0.09$ to +0.62 mag.
All standard stars were analyzed using the PHOTOMETRY task in DAOPHOTII
(Stetson 1995). With the results of aperture photometry,
a growth curve analysis was performed using DAOGROW (Stetson 1990)
to obtain integrated magnitudes.
The first-order $K$ passband extinction term $k_k$ is
0.077 mag air mass$^{-1}$ and the first-order color extinction terms
$k^{\prime}_{jk}$ and $k^{\prime}_{jh}$ are $-$0.003 and 0.039
mag air mass$^{-1}$, respectively.

To calibrate the photometry, the photometric transformations were assumed
to have the following form.
\begin{eqnarray}
\label{eqn:trans}
K_\mathrm{UKIRT} &=& k_0 + \zeta_K, \nonumber \\
(J-K)_\mathrm{UKIRT} &=& \mu(j-k)_0 + \zeta_{J-K}, \nonumber \\
(J-H)_\mathrm{UKIRT} &=& \epsilon(j-h)_0 + \zeta_{J-H}.
\end{eqnarray}
The transformation coefficients are $\zeta_K = -3.285 \pm 0.009$,
$\mu = 1.024 \pm 0.018$, $\zeta_{J-K} = 0.626 \pm 0.009$,
$\epsilon = 0.983 \pm 0.016$, and $\zeta_{J-H} = 0.115 \pm 0.004$
in the UKIRT $JHK$ system, where the errors are those of the mean.

Point-spread function (PSF) photometry for all NGC~6791 frames was
performed using DAOPHOTII and ALLSTAR/ALLFRAME
(Stetson 1987, 1994, 1995; Turner 1995).
We applied apperture corrections, air-mass corrections,
and the above photometric transformations
to the PSF magnitudes returned from our ALLFRAME run.
Finally, we converted our UKIRT $JHK$ magnitudes to the CIT $JHK$ magnitudes
using the following transformation equations given by
Casali \& Hawarden (1992):
\begin{eqnarray}
\label{eqn:ukirt2cit}
K_\mathrm{CIT} &=& K_\mathrm{UKIRT} - 0.018(J-K)_\mathrm{UKIRT}, \nonumber \\
(J-K)_\mathrm{CIT} &=& 0.936(J-K)_\mathrm{UKIRT}, \nonumber \\
(H-K)_\mathrm{CIT} &=& 0.960(H-K)_\mathrm{UKIRT}.
\end{eqnarray}

\section{RESULTS}

\subsection{The Infrared Color-magnitude diagram}

Figure~\ref{fig:fig1} shows the results of our
observations. The red giant branch, red horizontal branch, subgiant
branch, and even main sequence turn-off are clearly discernible.
Data for individual stars are given in Table~\ref{tab:mayallresults}.
Positions for these stars are taken from SBG2003.

\subsection{Comparison of 2MASS Photometry with Our Results}

Excluding stars near the edges of our frames, we are able
to match 149 stars in the 2MASS survey with stars for which
we have been able to measure $JHK$ magnitudes and colors. This
includes only 2MASS stars with the highest quality
photometry (``AAA" photometry quality flags). We have used
the recommended transformations between the 2MASS and CIT
photometric systems (Carpenter 2001), and
in Figures~\ref{fig:fig2}, \ref{fig:fig3},
and \ref{fig:fig4} we show the differences (in the sense
our results minus those from 2MASS) between the $K$ magnitudes
and $J-K$ and $H-K$ colors for the brighter stars,
those with $K \leq\ 13.6$ mag. The weighted mean difference in
$K$ is $-0.014 \pm 0.005$ mag, with no dependence on $K$ magnitude.
Likewise, there is no significant correlation in the differences
in the $J-K$ colors with the colors themselves, and
the weighted average difference between our values and those
from 2MASS is only $+0.007 \pm 0.005$ mag. In the case
of $H-K$, however, there is a correlation
between $\Delta$($H-K$) and $H-K$ (the correlation coefficient
is 0.68). The overall effect is modest, and the
mean difference in $H-K$ is only $-0.021 \pm 0.007$ mag.
We do not believe the problem arises in the $H$ filter we
employed, nor in our photometric data reductions since
the transformations appear to be good between the instrumental
system and the UKIRT standard stars. However, we note that
the equations derived by Carpenter (2001) for transforming
UKIRT and CIT photometry into the 2MASS system may be employed
to derive the following CIT-to-UKIRT relations:
\begin{eqnarray}
\label{eqn:carpenter}
K_\mathrm{CIT} &=& K_\mathrm{UKIRT} + 0.004(J-K)_\mathrm{UKIRT} - 0.026, \nonumber \\
(J-K)_\mathrm{CIT} &=& 1.012(J-K)_\mathrm{UKIRT} + 0.001, \nonumber \\
(H-K)_\mathrm{CIT} &=& 1.035(H-K)_\mathrm{UKIRT} + 0.045.
\end{eqnarray}
These do not agree particularly well with the relations derived
by Casali \& Hawarden (1992; Equation~\ref{eqn:trans}). The
relatively good agreement between our $K$ and $J-K$ results and those
from 2MASS, and the disagreement for the $H-K$ results suggests that
the transformations between the various photometric systems remain
a source of some uncertainties.

\subsection{Matching to Optical Photometry}

The very extensive $BVI$ photometry of SBG2003 was employed to obtain
optical photometry for our program stars. The transformation
of (x,y) positions in the superposed image of the cluster
began with identification of sixteen of the brighter
stars in the 2MASS survey, and a pair of third-order polynomial equations
to transform (x,y) to right ascension and declination.
The rms error of the fits were
of order 0.2\arcsec. The superposition of the five fields was not
perfect, and only about half of the 150 or so 2MASS stars were
identified by this process. A similar procedure was employed for
sixteen of the brighter
remaining 2MASS stars. The rms error in positions was roughly 0.1\arcsec\
for this case. We thus were forced to employ two possible coordinate
transformations for all of our program stars.

We then applied both transformations
to compute the estimated right ascensions and declinations for all of our
program stars. A match to a star from SBG2003 was declared only
if the separation was less than 3\arcsec. In the case of multiple
matches, only the nearest star was selected for each transformation.

Nonetheless, some of our program stars had matches using both transformations.
To select the star that provides a better match, we then compared $J-K$
color indices from our work with the $V$ magnitude from SBG2003 and
the $K$ magnitude from our work. A simple linear relation was fit to
the $J-K$ vs.\ $V-K$ color indices, and 3$\sigma$ outliers were rejected.
A new iteration was undertaken with the remaining stars, and multiple
iterations were employed until the relation showed no significant
change from one iteration to the next. At that point the
empirical $J-K$ vs.\
$V-K$ relation was employed to select the better match when both
position transformations found matches. We then eliminated
stars that deviated by more than 1.0 mag in $V-K$ from that predicted
by our $J-K$ photometry.

Our final results, given in Table~\ref{tab:mayallresults}, include
the positions for each star from RBG2003, the $V$, $B-V$, and $V-I$
photometry, also from SBG2003, and our own results, $J$, $H$,$K$,
as well as $J-K$ and $V-K$, and their associated errors.

\section{DISCUSSION}

\subsection{The Slope of the Red Giant Branch}

\subsubsection{Calibration of the relation between the slope and [Fe/H]}

The slope and color of the red giant branch have both been used for
decades to infer the mean metallicity of globular clusters from
optical color-magnitude diagrams. Zinn \& West (1984) and Zinn (1985)
discussed the earlier work thoroughly, and Sarajedini (1994) showed
how the reddening-insensitive slope and the reddening-dependent color
of a $BV$ color-magnitude diagram could be used to infer the
cluster metallicity and the reddening simulataneously. Sarajedini \&
Layden (1997) extended this to $VI$ color-magnitude diagrams.
Kuchinski et al.\ (1995)
and Kuchinski \& Frogel (1995) demonstrated that the slope of the
red giant branch in a $K$ vs.\ $J-K$ color-magnitude diagram could
also be employed as a metallicity indicator, but also argued
that the color of the red giant branch is not particularly sensitive
to reddening for metallicities in the range $-1.0 \leq$\ [Fe/H] $\leq\ -0.3$,
diminishing the possibility of measuring both [Fe/H] and E($J-K$)
simultaneously.

Kuchinski et al.\ (1995) recommended simple linear fits to determine
the slope since, unlike optical photometry, the red giant branches
do not show any significant curvature above the level of the horizontal
branch. They recommended using stars lying between 0.6 and 5.1 magnitudes
above the level of the red horizontal branch, the bright limit being
set by the desire to avoid possible photometric variable stars near
the luminous tip of the red giant branch. Valenti et al.\ (2004b)
and VFO have re-calibrated
the relationship between the slope, defined by $\Delta$($J-K$)/$\Delta K$,
and [Fe/H]. They noted that reliance on the magnitude level of the
horizontal branch to define a reference point compromises the
method prescribed by Kuchinski et al.\ (1995) since some
clusters do not have red horizontal branch stars. Since the
``horizontal branch" is not horizontal in a $K$ vs.\ $J-K$ diagram,
clusters with predominantly blue horizontal branches would be more
difficult to exploit within the above magnitude limit prescription.
Valenti et al.\ (2004b) and VFO therefore recommended
that the slope be simply defined using stars 0.5 and 5.0 magnitudes fainter than
the most luminous red giant branch star. We point out the obvious
fact that in the case of a linear fit to a linear set of data will
yield exactly the same answer, so either set of measured slopes
should, in principle, suit equally well. We show below, however,
that the RGB slope does \underline{not} appear to be linear.

Both Valenti et al. (2004b) and VFO also compared the
slope with [M/H], the ``mean metallicity" of the cluster. This is meant
to include the contributions of elements other than those of the iron
peak, the abundance of light elements in particular. The logic of such
a calibration is that model isochrones depend on the heavy element
mass fraction, $Z$, and results are also often available for different
mixes of the ``$\alpha$" elements (O, Mg, Ca, Si, and in some cases Ti).
Indeed, isochrones that are distinguished only by $Z$ and not by
[$\alpha$/Fe] may still be compared using an expression relating them
to [Fe/H] (see Straniero \& Chieffi 1991 and Salaris, Chieffi, \& Straniero
1993 for details). The most inclusive calibration between [M/H] and
the slope of the red giant branch is that given by VFO. We are
reluctant to employ such a calibration for NGC 6791 or, indeed,
the most heavily extinguished globular clusters for which observations
are now underway, because systematic differences between
[$\alpha$/Fe] at fixed [Fe/H] do exist to a small degree among field
halo stars (King 1997; Carney et al.\ 1997; Nissen \& Schuster 1997),
in a few globular clusters (Brown, Wallerstein, \& Zucker 1997; Cohen 2004), and
in stars in the Galactic bulge (McWilliam \&
Rich 1994). The differences are even stronger in
neighboring dwarf galaxies with predominantly old populations
such as Draco and Ursa Minor (Shetrone, C\^{o}t\'{e}, \&
Sargent 2001) and Sculptor (Shetrone et al.\ 2003), where
low [$\alpha$/Fe] ratios are found even at very low
metallicities.
Comparisons with model isochrones using [M/H] or [Fe/H] and
variable [$\alpha$/Fe] values
are useful in an exploratory sense, as we do below, but the most reliable
comparisons can be made only once high-resolution spectroscopy has
provided as detailed an elemental abundance profile as possible.

VFO also derived a relationship between the slope of the red giant
branch and [Fe/H], and we reproduce their equation A41 here:
\begin{equation}
\label{eq:fecg97}
[Fe/H]_{CG97} = -22.21(slope_{RGB}) - 2.80.
\end{equation}
Note that the metallicity scale they employed was that of Carretta \& Gratton
(1997), which differs slightly from that employed by Zinn (1985). VFO
also noted that this relation disagrees significantly with that
derived by Ivanov \& Borissova (2002). VFO attributed the difference
to the less extensive set of data available to Ivanov \& Borissova (2002)
and to an inaccurate measurement of the red giant branch slope for
47~Tuc, their most metal-rich cluster.

We employ here the [Fe/H] metallicity scale adopted by Zinn (1985).
While yet another calibration
of the relationship between the red giant branch slope and [Fe/H] may
be of limited value, we believe such a calibration has merit.
The Carretta \& Gratton (1997)
metallicity scale is not as widely applicable to other clusters as
those of Zinn (1985). Indeed, excluding the multi-metallicity
cluster $\omega$~Cen, use of the results from Zinn (1985)
increases the sample size from 23 to 28 clusters,
and is capable of growing much larger still using an internally consistent
metallicity scale as new $JHK$ observations of clusters become available.
We collect in Table~\ref{tab:rgbslopesvsfe} the available data, relying primarily on VFO
for the data for the red giant branch slopes, and the [Fe/H] values
from Zinn (1985). Use of an ordinary least squares bissector to
the data, assuming $\sigma$([Fe/H]) $\approx\ 0.2$ for most clusters,
results in:
\begin{equation}
\label{eq:fez85}
[Fe/H]_{Zinn} = -27.95 (\pm 2.26)(slope_{RGB}) - 3.35 (\pm 0.19).
\end{equation}
Figure~\ref{fig:fig5} shows our results.
Please note that the open circle represents
our own results (Lee, Carney, \& Balachandran 2004) for Palomar~6,
a cluster so heavily extinguished that it is most readily studied
photometrically {\em and} spectroscopically only at infrared wavelengths.

\subsubsection{Results for NGC 6791}

As discussed in the Introduction, NGC~6791 is thought to have
a very high metallicity, [Fe/H] $\approx\ +0.3$. Extrapolation
of Equations~\ref{eq:fecg97} and \ref{eq:fez85} to such a
metallicity predict red giant branch slopes of $-0.14$ and
$-0.13$, respectively. If the uncertainty in the slope
of either relation is about 8\%, as implied in Equation~\ref{eq:fez85},
then these predicted slopes have uncertainties of about $\pm 0.01$.

NGC~6791 is a relatively sparse cluster, 
and coupled with the relatively small area we have
surveyed, there are too few stars at or
above the magnitude level of the horizontal branch to employ the
full 4 to 5 magnitude range recommended by either Kuchinski et al.\ (1995)
or VFO. But only the precision of the slope is compromised by a
more limited range, not the value itself, {\em so long as the data are
indeed well-defined by a linear fit}. Because our results for the
slope are not consistent with the predictions, we derive them
here using three different approaches. 

First, to increase the
number of stars and, we hope, improve the precision of our
measure, we extend the fitting region below the horizontal branch.
Figure~\ref{fig:fig6} shows the
resultant slope and the stars employed to
determine the fit (filled circles). We have again employed an
ordinary least squares bissector fitting routine. The slope
in this case is $-0.088 \pm 0.005$, which differs considerably
from what is expected. In fact, both Equations~\ref{eq:fecg97} and
\ref{eq:fez85} then predict [Fe/H] = $-0.9 \pm 0.1$. 

Second, we restrict the measurement of the slope to only the
nine brightest red giants in
Figure~\ref{fig:fig6}, avoiding the problem of possible
non-linear behavior for stars fainter than the horizontal branch.
In this case,
the slope becomes $-0.092 \pm 0.013$,
implying [Fe/H] = $-0.8 \pm 0.3$. The implied very low metallicity
is confirmed. The dashed line shows roughly what the slope
should be if we use Equation~\ref{eq:fez85}, with the intersection
between the two relations forced to lie at the magnitude level
of the horizontal branch. It is very clear that our data yield
a red giant branch slope that is inconsistent with a metallicity
greater than solar, if Equations~\ref{eq:fecg97} or \ref{eq:fez85}
are extrapolated into higher metallicity regions than the
clusters that have been used to calibrate both relations.

Finally, we derive slope using the results from
the 2MASS survey, independent of our own
photometry. We do not expect different answers since, 
as we have seen, the 2MASS photomery is very consistent with
our data once minor zero point offsets are taken into account.
In Figures~\ref{fig:fig7},
\ref{fig:fig8}, and \ref{fig:fig9} we
show the results from 2MASS. Figure~\ref{fig:fig7}
shows the ``field" in the
vicinity of NGC 6791, and includes stars in four fields 15\arcmin\
north, south, east, and west of the cluster center. Each field has
a radius of 3\arcmin. These were selected to crudely sample possible
variations in interstellar reddening (see below) and provide a sample
size roughly consistent with that of Figure~\ref{fig:fig8},
which was centered
on the cluster and also has a radius of 3\arcmin. The filled circles
in Figure~\ref{fig:fig8} represent the 8
stars used for the fit, which yielded
a red giant branch slope of $-0.084 \pm 0.009$.
Figure~\ref{fig:fig9} repeats
the analysis, but includes a radius of 4\arcmin\ around the cluster
center, and a fit to the 11 stars marked as filled circles, which
yielded a slope of $-0.083 \pm 0.007$. Figures~\ref{fig:fig8}
and \ref{fig:fig9}, coupled
with Equations~\ref{eq:fecg97} and \ref{eq:fez85} all lead to
[Fe/H] estimates that range from $-0.9$ to $-1.0$, each with
an uncertainty of about $\pm 0.2$ dex. The 2MASS results thus
independently confirm that {\em the slope of the red giant branch in
NGC~6791 is not consistent with the calibrations of
Equation~\ref{eq:fecg97} and \ref{eq:fez85} and [Fe/H]
values anywhere near solar.} Something is wrong.

How can we reconcile the differences between the spectroscopic estimates
of the high metallicity for NGC~6791 and the much lower value
implied by the slope of the red giant branch? There are basically
six options.

\begin{enumerate}

\item Our photometry is in error. This also seems unlikely,
since our results agree so well with 2MASS, and we feel that our
observations were obtained under excellent observing conditions, and
optimized to remove any possible detector non-linearities.

\item The cluster's metallicity is, in fact, very low. This
likewise seems implausible since the spectroscopic evidence for
such a high metallicity is compelling.

\item NGC~6791 may mimic a lower metallicity cluster if the
dominant electron donors in the stellar atmospheres are
underabundant relative to the calibrating clusters. This
would include the $\alpha$ elements such as Ca and Mg, in
particular. This is in apparent conflict with the results
of Peterson \& Green (1998).

\item We have too few stars to reliably measure the slope of the
red giant branch. While the number of stars is small, the slope
does appear to be well-measured.

\item The slope of the red giant branch is a linear function
of [Fe/H], as indicated in Figure~\ref{fig:fig5},
but only if [Fe/H] $\leq\ -0.3$.

\item Finally, the claim that the slope of the red giant branch
is linear in $K$ vs.\ $J-K$ may be incorrect.

\end{enumerate}

We explore the explore that latter two points using the isochrones
of Yi, Kim, \& Demarque (2003; hereafter YKD2003) and the 
color-temperature relations of Alonso et al.\ (1999).
Figure~\ref{fig:fig10} shows the predicted values of $M_{K}$ vs.\
$J-K$ for [Fe/H] values ranging from $-0.8$ to +0.3. Overall, it
appears that the color of the red giant branch continues to depend
on [Fe/H] up to the metallicity comparable to that of NGC~6791.
However, the slope of the red giant branch at any
fixed metallicity is {\em not} predicted to
be linear above $M_{K}$ = $-1.5$, a value typical of the red
horizontal branch or red clump stars. Curvature in the $M_{K}$ vs.\
$J-K$ plane is clearly present at each metallicity shown. While
linear approximations of the slope may yet prove useful, the
Figure shows that any such measure must employ consistently
defined and employed reference points. Nontheless, the
the slope of the red giant branch does appear to still be a good
reddening-insensitive metallicity
indicator, although it appears that a new calibration will be required
so that it may be applied to sparse clusters, such
as NGC~6791, whose red giants
do not extend more than about two magnitudes above the red
horizontal branch/red clump domain.

\subsection{Cluster Reddening}

NGC~6791 lies at a relatively high Galactic latitude ($b \approx 10.9$ deg),
so the reddening is not expected to be high, as discussed in
detail by SBG2003. The compilation of Twarog, Ashman, \&
Anthony-Twarog (1997) suggests E($B-V$) is 0.15 mag, while SBG2003
adopted 0.09 mag, and Montgomery et al.\ (1994) found a value of
$0.10 \pm 0.02$ mag using field stars in the cluster's vicinity.

Our photometry may be used to
explore this issue, following the suggestions of
Kuchinski et al.\ (1995)
and Kuchinski \& Frogel (1995) that
the mean color of the red horizontal
branch for globular clusters with $-1.0 \leq$\ [Fe/H] $\leq\ -0.3$
is insensitive to metallicity.

We revisit this issue here. We confine our attention to only the oldest
clusters: those with estimated ages exceeding 2.5 $\times\ 10^{9}$ years.
This includes only three open clusters (Be~39, NGC~6791, and NGC~6819)
from Grocholski \& Sarajedini (2002), plus their globular clusters 47~Tuc,
NGC~362, and the globular clusters NGC~5927, NGC~6624, NGC~6637, NGC~6712,
and NGC~6838 (M71) from Kuchinski \& Frogel (1995). (We exclude the most
heavily extinguished clusters Liller~1, Terzan~2, and NGC~6440.) Before
presenting our results, we felt it necessary to repeat the color
estimations done by Grocholski \& Sarajedini (2002), since it appeared
from their figures that their results may have been compromised by
including stars on the red giant branch. We show in Figures~\ref{fig:fig11}
and \ref{fig:fig12} the 2MASS $K$ vs.\ $J-K$ data for these clusters,
with the red horizontal branch/red clump stars identified as open circles.
The mean colors were computed and transformed to the CIT photometric
system (Carpenter 2001), and our results are given in Table~\ref{tab:rhb}.

Figure~\ref{fig:fig13} somewhat confirms the speculation of
Kuchinski \& Frogel (1995),
which is hardly suprising since we are employing their data for the
clusters with $-1.0 \leq$\ [Fe/H] $\leq\ -0.3$. But the Figure also
shows that inclusion of more metal-poor and more metal-rich clusters
shows that there is indeed a fairly strong metallicity sensitivity!
If we fit the data using an ordinary least squares bissector, we find
\begin{equation}
\label{eq:jkvsfe}
<J-K>_{0} = (0.170 \pm 0.026) [Fe/H] + (0.596 \pm 0.016).
\end{equation}

The fifteen stars identified as red horizontal branch stars in
Figure~\ref{fig:fig6} yield $<J-K>$ = $0.73 \pm 0.01$.
If we accept that [Fe/H] = $+0.4 \pm 0.1$ (Peterson \& Green 1998),
we find that E($J-K$) = $0.075 \pm 0.02$. 
Employing the interstellar extinction law measured by
Rieke \& Lebofsky (1985), we find E($B-V$) = $0.14 \pm 0.04$.
This is consistent with previous cluster estimates. Indeed, the estimate
of Montgomery et al.\ (1994) of $0.10 \pm 0.02$ would, notionally,
support a metallicity for the cluster a bit higher than +0.4.

\subsection{Cluster Distance}

The luminosity of the horizontal branch has been an extremely useful
(and disputed) tool for the estimation of cluster distances for
decades. Since the infrared magnitudes are less vulnerable to
reddening and extinction, $M_{K}$(RHB) has attracted some special
attention. Kuchinski et al.\ (1995) and Kuchinski \& Frogel (1995)
helped pioneer this area of study as well, noting that the intermediate
metallicity clusters had comparable $M_{K}$(RHB) values. Alves (2000)
combined $Hipparcos$ parallaxes, $K$-band photometry, and spectroscopic
metallicity estimates for over two hundred field red clump stars, almost
all with $-0.6 \leq$\ [Fe/H] $\leq\ +0.1$. The peak in the luminosity
distribution was found to lie at $M_{K} \approx\ -1.61 \pm 0.03$, and
Alves (2000) argued that there is no sensitivity to metallicity,
at least over the range studied, making the tool particularly powerful
for metal-rich clusters, and supporting the earlier results of
Kuchinski et al.\ (1995) and Kuchinski \& Frogel (1995). We
note that apparently not all of the stars studied by Alves (2000)
are horizontal branch or red clump stars since the dispersion in
the derived $M_{K}$ value is much too large. But the median value
appears reliable, and, in fact, Grocholski \& Sarajedini (2002)
derived a remarkably similar value, $M_{K}$ = $-1.62 \pm 0.06$,
using open clusters. Age may also play a role, and ages for
the field stars employed by Alves (2000) are unknown.
Grocholski \&
Sarajedini (2002) explored this issue as well,
employing a consistent set of distance, reddening, and metallicity
estimation procedures for a sample of open and globular
clusters. They found that
clusters with ages greater than about 2~Gyrs did not
show significant variation in $M_{K}$(RHB) with age, but that
there is some sensitivity to metallicity.
(Conversely, younger clusters' red clump stars' luminosities were
not so dependent on metallicity, but were sensitive to age differences.)
Finally, Pietrzy\'{n}ski, \& Udalski (2003) compared $K$-band red clump
luminosities derived for stars in four Local Group galaxies with
distances estimated from the the red giant branch tip $I$-band magnitudes,
$V$-band RR Lyrae luminosities, and $K$-band cepheid observations. They
agreed with Gocholski \& Sarajedini (2000) that the $K$-band luminosity
of red clump stars has ``very little (if any) dependence on age over an
age range of about 2-8 Gyrs". They similarly found little sensitivity
to metallicity, although their metallicity range was roughly from
$-0.5$ down to $-1.8$, well out the regime of interest in terms
of a distance estimate for NGC~6791.

The value of $<K>$ for the fifteen red horizontal branch stars
in NGC~6791 is $11.50 \pm 0.03$, and differences in E($B-V$) of
0.05 mag alter the calculation of $<K_{0}>$ by only 0.01 mag.
Using the calibration of Alves (2000), we therefore derive
a distance modulus of ($m-M$)$_{0}$ = $13.07 \pm 0.04$.

\section{THE AGE OF NGC 6791}

The color index $V-K$ provides a very powerful indicator of
effective temperature in stars, and its relationship to
$T_{\rm eff}$ and [Fe/H] for dwarfs and giants has been
described by Alonso et al.\ (1996, 1999), respectively.
In brief, the wide wavelength span enables high sensitivity,
roughly 20~K per 0.01 mag, or better. The metallicity sensitivity
is also weak, at least until temperatures are cool enough
or metallicities are high enough
for TiO bands to cause significant absorption in the $V$ bandpass.

Figures~\ref{fig:fig14} and \ref{fig:fig15}
show the combined optical and infrared photometry for our sample
of stars using the $V-K$ color index and the $K$ and $V$ magnitudes,
respectively. The effects of TiO absorption in the $V$ bandpass
are apparent. In Figure~\ref{fig:fig14} the shape of
the cooler part of the red giant branch has a noticeable change
of slope, compared to the near-linear behavior seen in the
$K$ vs.\ $J-K$ diagram shown in Figure~\ref{fig:fig1}.
In Figure~\ref{fig:fig15}, where we compare $V$ vs.\ $V-K$,
the effect is even stronger. Optical color indices such as $B-V$
and $V-I$, with shorter wavelength baselines, coupled with the
$V$ bandpass, are therefore not as suitable for thoroughly
testing observations with the predictions of model isochrones.

We have relied on the grid of model isochrones computed by
YKD2003, and the transformation of luminosity into $M_{V}$ and
$M_{K}$ and effective temperature into $B-V$, $J-K$, and $V-K$
that they employed. Because our infrared data are on a slightly
different photometric system, we employed the transformations
recommended by Bessell \& Brett (1998). Because good empirical color-temperature
transformations exist only for solar-metallicity stars with
[$\alpha$/Fe] $\approx$\ 0.0, we chose to employ only
those isochrones with a solar mix of heavy elements.

Figure~\ref{fig:fig16} shows fits in the four
color-magnitude diagram planes for [Fe/H] = 0.0, using parameters selected
according to the criterion that a ``best fit" be obtained
in the $V$ vs.\ $B-V$ plane. Note that the implied reddening,
E($B-V$) = 0.26 mag, and distance modulus, ($m-M$)$_{0}$ = 12.73,
are not in good agreement with the estimates presented above.
Note also that while the match is very good in the optical,
the matches in the infrared are, frankly, unacceptable, due
primarily to the large reddening value we had to adopt.
(The unusual ``hook" behavior seen in the $V$ vs.\ $B-V$ plane
near the tip of the red giant branch arises from line blanketing
produced by molecular absorption, especially by TiO. The optical
spectra of cooler, more metal-rich stars are affected especially.
The Figure illustrates that near-infrared spectra and photometry
are affected much less strongly.)

Figure~\ref{fig:fig17} shows one of the best
matches for [Fe/H] = +0.3, E($B-V$) = 0.17 mag, ($m-M$)$_{0}$ = 12.93,
and an age of 8.0 Gyrs.
A slightly higher extinction (0.21 mag) leads
to ($m-M$)$_{0}$ = 13.00 mag, and an age of 6.0 Gyrs, but the fit
is slightly poorer. Increasing [Fe/H] to +0.4, the value
recommended by Peterson \& Green (1998), leads to Figure~\ref{fig:fig18},
with an optimum optical fit being obtained for E($B-V$) = 0.13 mag,
($m-M$)$_{0}$ = 12.96 and an age of 8 Gyrs. Now the agreement
with the estimated reddening and distance modulus is even better, and
the agreement for all the color-magnitude diagrams with the
isochrones is quite good.

The best match to the empirical reddening and distance estimates
and the best matches to color-magnitude diagrams is shown in
Figure~\ref{fig:fig19}, for [Fe/H] = +0.5, E($B-V$) = 0.11 mag,
($m-M$)$_{0}$ = 13.11 mag, and an age of 7.5 Gyrs.
The agreement
of the reddening and the distance modulus to the empirically-derived
values discussed in Sections~5.2 and 5.3 is excellent. Adopting
[Fe/H] = +0.5 leads to empirically-calibrated
values for E($B-V$) of 0.09 mag and ($m-M$)$_{0}$ = 13.07 mag.
We note that this metallicity is slightly higher than derived by
Peterson \& Green (1998), but is within the uncertainty, and
may be accounted for by our use of [$\alpha$/Fe] = 0.0, while
a value of +0.1 may be more appropriate.

While there are uncertainties in all of our estimates, including
the reddening, the distance modulus, and in the procedure of
fitting model isochrones to the optical and infrared color-magnitude
diagrams, we are very encouraged by the consistency of the results
for the relatively high [Fe/H] values of +0.4 and +0.5, from which
we conclude that the isochrones are most consistent with a cluster
age of roughly 8 Gyrs.

\section{SUMMARY}

We have obtained relatively deep $JHK$ photometry of the very old,
very metal-rich open cluster NGC~6791. We have derived a revised
relationship between the slope of the red giant branch and the
metallicities of clusters using the Zinn (1985) calibrations.
However, the slope of the RGB in NGC~6791 yields an unrealistically
low metallicity because the cluster RGB is not well populated and
the RGB itself is not expected to be linear, even
in the $K$ vs.\ $J-K$ plane.
The mean color of the red horizontal branch
stars implies E($B-V$) = $0.14 \pm 0.04$ mag, while the mean $K$
magnitude is consistent with ($m-M$)$_{0}$ $\approx\ 13.07$. Both
of these values produce very good agreement between optical and infrared
color-magnitude diagrams for [Fe/H] values of +0.4 or +0.5.
The optimal matches are found for a cluster age of roughly 8 Gyrs. Lower
metallicities yield poorer agreements between the color-magnitude
diagrams and model isochrones.

JWL thanks Y.\ -C.\ Kim for his kind discussion and acknowledges
the financial support of the Korea Science and Engineering Foundation 
(KOSEF) to the Astrophysical Research Center for the Structure 
and Evolution of the Cosmos (ARCSEC).
BWC acknolwedges the financial support of the National Science Foundation
to the University of North Carolina through grants AST-9619381,
AST-9988156, and AST-0305431.

\clearpage

\begin{deluxetable}{llrrrrrrrr}
\footnotesize
\tablecaption{Results of $JHK$ Photometry \& Matching to Optical Photometry
\label{tab:mayallresults}}
\tablenum{1}
\tablewidth{0pc}
\tablehead{
\colhead{Id} &
\colhead{Right ascension} &
\colhead{$V$} &
\colhead{$B-V$} &
\colhead{$V-I$} &
\colhead{$J$} &
\colhead{$H$} &
\colhead{$K$} &
\colhead{$J-K$} &
\colhead{$V-K$} \\
\colhead{} &
\colhead{Declination} &
\colhead{$\sigma$} &
\colhead{$\sigma$} &
\colhead{$\sigma$} &
\colhead{$\sigma$} &
\colhead{$\sigma$} &
\colhead{$\sigma$} &
\colhead{$\sigma$} &
\colhead{$\sigma$} }
\startdata
10099 & 19 20 58.65 & 12.155 &  2.348 &  0.569 &  9.658 &  9.064 &  8.668 &  0.990 &  3.487\\
  & 37 47 40.8  &  0.047 &  0.224 &  0.086 &  0.075 & 0.058 &  0.051 &  0.091 &  0.070\\
 8904 & 19 20 55.11 & 13.862 &  1.633 &  1.970 & 10.582 &  9.780 &  9.598 &  0.984 &  4.264\\
  & 37 47 16.5  &  0.000 &  0.002 &  0.001 &  0.029 & 0.028 &  0.018 &  0.034 &  0.018\\
 8266 & 19 20 53.39 & 13.741 &  1.616 &  1.779 & 10.775 & 10.024 &  9.770 &  1.005 &  3.971\\
  & 37 48 28.4  &  0.000 &  0.002 &  0.001 &  0.058 & 0.054 &  0.037 &  0.069 &  0.037\\
11814 & 19 21  4.27 & 13.849 &  1.575 &  1.658 & 11.060 & 10.304 & 10.128 &  0.932 &  3.721\\
  & 37 47 18.9  &  0.001 &  0.001 &  0.001 &  0.049 & 0.049 &  0.029 &  0.057 &  0.029\\
 7972 & 19 20 52.60 & 14.136 &  1.593 &  1.763 & 11.160 & 10.445 & 10.208 &  0.952 &  3.928\\
  & 37 44 28.5  &  0.000 &  0.001 &  0.001 &  0.050 & 0.047 &  0.032 &  0.059 &  0.032\\
 7063 & 19 20 50.04 & 13.061 &  1.081 &  1.249 & 10.927 & 10.356 & 10.259 &  0.668 &  2.802\\
  & 37 47 28.4  &  0.004 &  0.006 &  0.005 &  0.031 & 0.028 &  0.018 &  0.036 &  0.019\\
 5342 & 19 20 44.86 & 14.138 &  1.569 &  1.749 & 11.203 & 10.463 & 10.298 &  0.905 &  3.840\\
  & 37 46 21.7  &  0.000 &  0.001 &  0.001 &  0.046 & 0.043 &  0.029 &  0.054 &  0.029\\
 7328 & 19 20 50.78 & 13.599 &  0.869 &  1.259 & 11.618 & 11.119 & 11.006 &  0.612 &  2.593\\
  & 37 46 33.8  &  0.000 &  0.003 &  0.001 &  0.042 & 0.036 &  0.025 &  0.049 &  0.025\\
\enddata
\end{deluxetable}

\clearpage

\begin{deluxetable}{lrcccc}
\footnotesize
\tablecaption{Calibrating Clusters for the Relation Between [Fe/H]
and the Slope of the Red Giant Branch \label{tab:rgbslopesvsfe}}
\tablenum{2}
\tablewidth{0pc}
\tablehead{
\colhead{Cluster} &
\colhead{[Fe/H]\tablenotemark{a}} &
\colhead{Slope} &
\colhead{$\sigma$} &
\colhead{Reference\tablenotemark{b}} }
\startdata
NGC 104 (47 Tuc) & $-0.71$ & $-0.110$ & 0.002 & 1 \\
NGC 288 & $-1.40$ & $-0.071$ & 0.004 & 1 \\
NGC 362 & $-1.28$ & $-0.074$ & 0.003 & 1 \\
NGC 4590 (M68) & $-2.09$ & $-0.048$ & 0.003 & 1 \\
NGC 5272 (M3) & $-1.66$ & $-0.071$ & 0.003 & 1 \\
NGC 5904 (M5) & $-1.40$ & $-0.082$ & 0.004 & 1 \\
NGC 5927 & $-0.30$ & $-0.112$ & 0.005 & 2 \\
NGC 6121 (M4) & $-1.28$ & $-0.079$ & 0.009 & 1 \\
NGC 6171 (M107) & $-0.99$ & $-0.101$ & 0.005 & 1 \\
NGC 6205 (M13) & $-1.65$ & $-0.065$ & 0.002 & 1 \\
NGC 6254 (M10) & $-1.60$ & $-0.048$ & 0.005 & 1 \\
NGC 6341 (M92) & $-2.24$ & $-0.046$ & 0.003 & 1 \\
NGC 6342 & $-0.66$ & $-0.102$ & 0.003 & 1 \\
Terzan 2 & $-0.47$ & $-0.107$ & 0.006 & 2 \\
NGC 6380 & $-0.99$ & $-0.094$ & 0.003 & 1 \\
Palomar 6 & $-0.8$ & $-0.075$ & 0.004 & 3 \\
NGC 6440  & $-0.26$ & $-0.093$ & 0.005 & 1 \\
NGC 6441 & $-0.53$ & $-0.092$ & 0.005 & 1 \\
NGC 6528  & $+0.12$ & $-0.114$ & 0.002 & 1 \\
NGC 6553 & $-0.29$ & $-0.092$ & 0.002 & 1 \\
NGC 6624 & $-0.35$ & $-0.100$ & 0.005 & 4 \\
NGC 6637 (M69) & $-0.59$ & $-0.092$ & 0.002 & 1 \\
NGC 6712 & $-1.01$ & $-0.091$ & 0.006 & 2 \\
NGC 6752 & $-1.54$ & $-0.048$ & 0.003 & 1 \\
NGC 6809 (M55) & $-1.82$ & $-0.049$ & 0.003 & 1 \\
NGC 6838 (M71) & $-0.58$ & $-0.110$ & 0.008 & 2 \\
NGC 7078 (M15) & $-2.15$ & $-0.044$ & 0.003 & 1 \\
NGC 7099 (M30) & $-2.13$ & $-0.044$ & 0.004 & 1 \\
\enddata
\tablenotetext{a}{[Fe/H] values are from Zinn (1985), except
for that of Palomar~6, which is taken from Lee et al.\ (2004).}
\tablenotetext{b}{(1) Valenti et al.\ (2004a); (2) Kuchinski
et al.\ (1995); (3) Lee \& Carney (2002); (4) Kuchinski \& Frogel (1995)}
\end{deluxetable}

\clearpage

\begin{deluxetable}{lrrrrrrrr}
\footnotesize
\tablecaption{Mean Magnitudes and Colors of Clusters' Red Horizontal
Branches or Red Clumps \label{tab:rhb}}
\tablenum{3}
\tablewidth{0pc}
\tablehead{
\colhead{Cluster} &
\colhead{E($B-V$)} &
\colhead{[Fe/H]} &
\colhead{$\sigma$} &
\colhead{log age} &
\colhead{$M_{K}$(RHB)} &
\colhead{$\sigma$} &
\colhead{$<$($J-K$)$>_{0}$\tablenotemark{a}} &
\colhead{$\sigma$} }
\startdata
\multicolumn{9}{c}{} \\
\multicolumn{9}{c}{Old Open Clusters\tablenotemark{b}} \\
\multicolumn{9}{c}{} \\
Be~39 &    0.11 & $-0.18$ & 0.03 & 9.88 & $-1.64$ & 0.12 & 0.605 & 0.02 \\
M67 &   0.04 &  0.000 & 0.09 & 9.60 & $-1.71$ & 0.108 & 0.600 & 0.01 \\
NGC~6819 &  0.16 & 0.07 & 0.04 & 9.42 & $-1.69$ & 0.136 & 0.56 & 0.02 \\
\multicolumn{9}{c}{} \\
\multicolumn{9}{c}{Globular Clusters\tablenotemark{b}} \\
\multicolumn{9}{c}{} \\
47~Tuc &   0.04 & $-0.70$ & 0.07 & 10.08 & $-1.364$ & 0.211 & 0.50 & 0.02 \\
\multicolumn{9}{c}{} \\
\multicolumn{9}{c}{Globular Clusters\tablenotemark{c}} \\
\multicolumn{9}{c}{} \\
NGC~362 &  0.05 & $-1.15$ & 0.06 & 10.08 & $-0.83$ & 0.241 & 0.39 & 0.04 \\
NGC~5927 &  0.45 & $-0.31$ & 0.1  & 10.08 & $-1.28$ & 0.2 & 0.49 & 0.02 \\
NGC~6624 &  0.25 & $-0.37$ & 0.10 & 10.08 & $-1.17$ & 0.2 & 0.50 & 0.02 \\
NGC~6637 &  0.18 & $-0.59$ & 0.10 & 10.08 & $-1.18$ & 0.2 & 0.49 & 0.02 \\
NGC~6712 &  0.39 & $-1.01$ & 0.10 & 10.08 & $-1.16$ & 0.2 & 0.44 & 0.02 \\
NGC~6838 & 0.27 & $ -0.58$ & 0.10 & 10.08 & $-1.03$ & 0.2 & 0.47 & 0.02 \\
\enddata
\tablenotetext{a}{All photometry is given in the CIT system}
\tablenotetext{b}{From Kuchinski \& Frogel (1995)}
\tablenotetext{c}{From Grocholski \& Sarajedini (2002)}
\end{deluxetable}

\clearpage

\section*{FIGURE CAPTIONS}

\figcaption[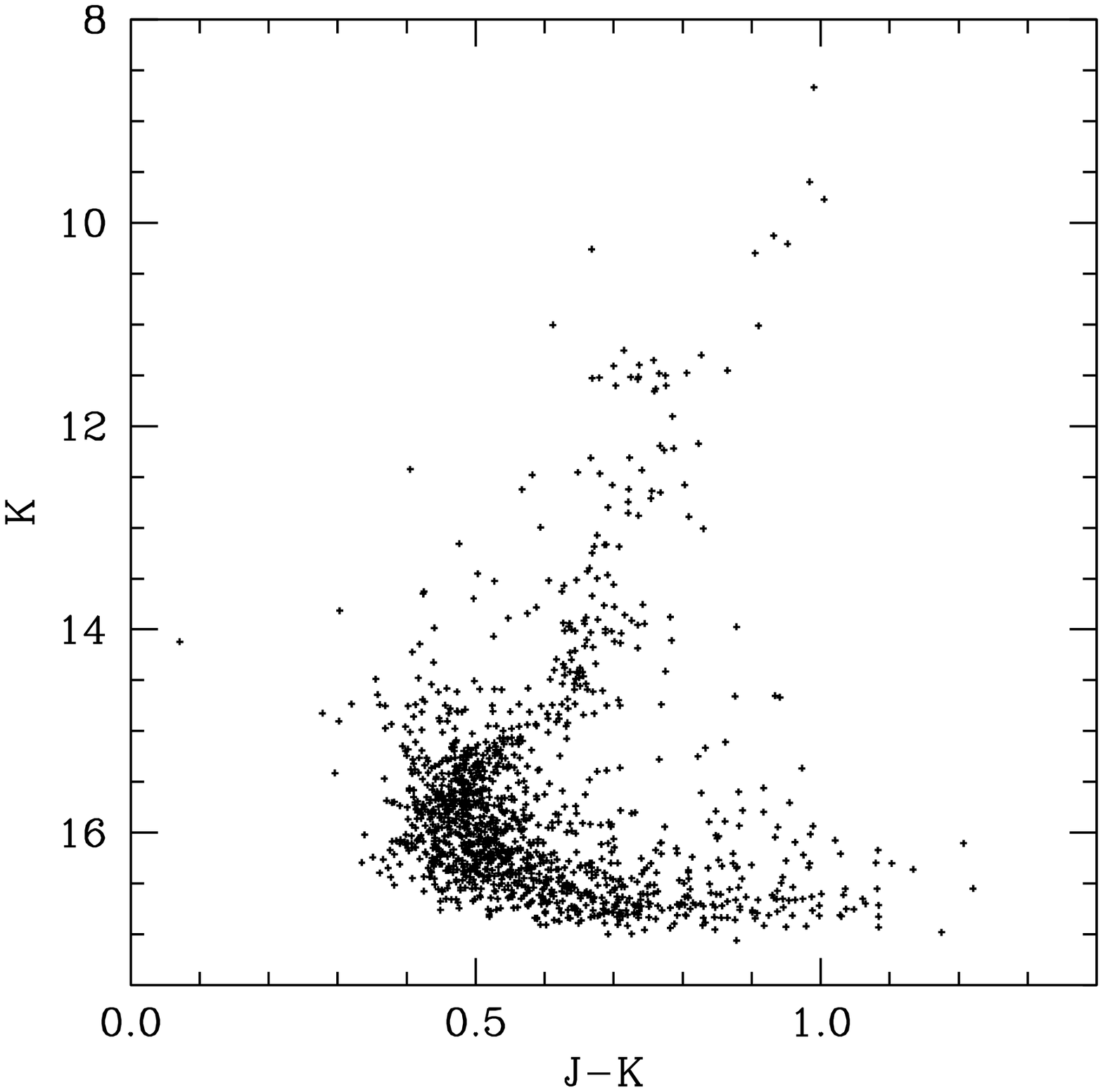]{The $K$ vs.\ $J-K$ diagram
resulting from our observations. The photometry is on the
``CIT" photometric system. \label{fig:fig1}}

\figcaption[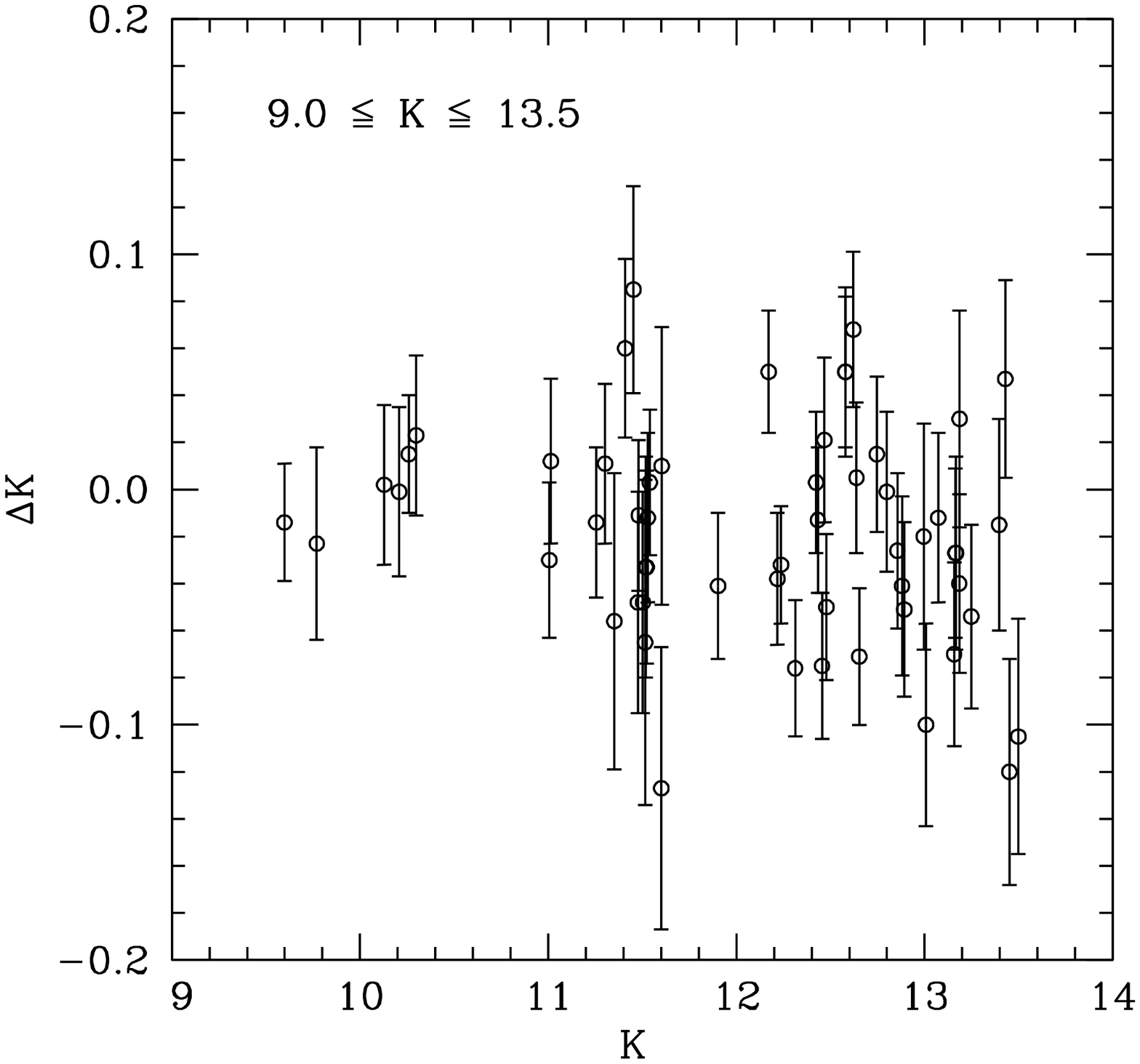]{The differences between our $K$ magnitude results
minus those from the 2MASS survey, after transformation to
the CIT photometric system. Only stars with highest quality
photometry and in a limited range of $K$ magnitudes are shown.
\label{fig:fig2}}

\figcaption[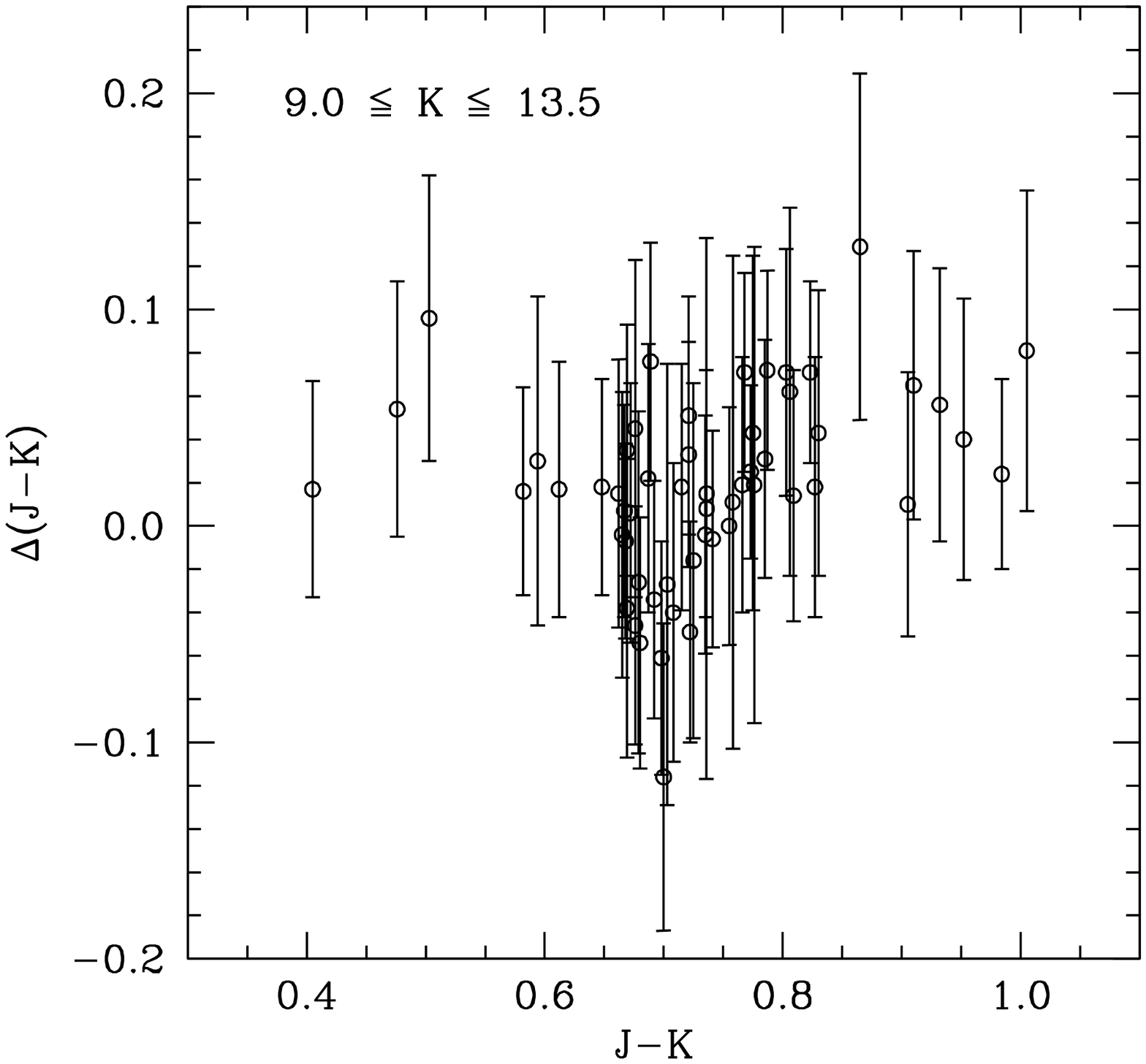]{The differences between our $J-K$ results
minus those from the 2MASS survey, after transformation to
the CIT photometric system. Only stars with highest quality
photometry and in a limited range of $K$ magnitudes are shown.
\label{fig:fig3}}

\figcaption[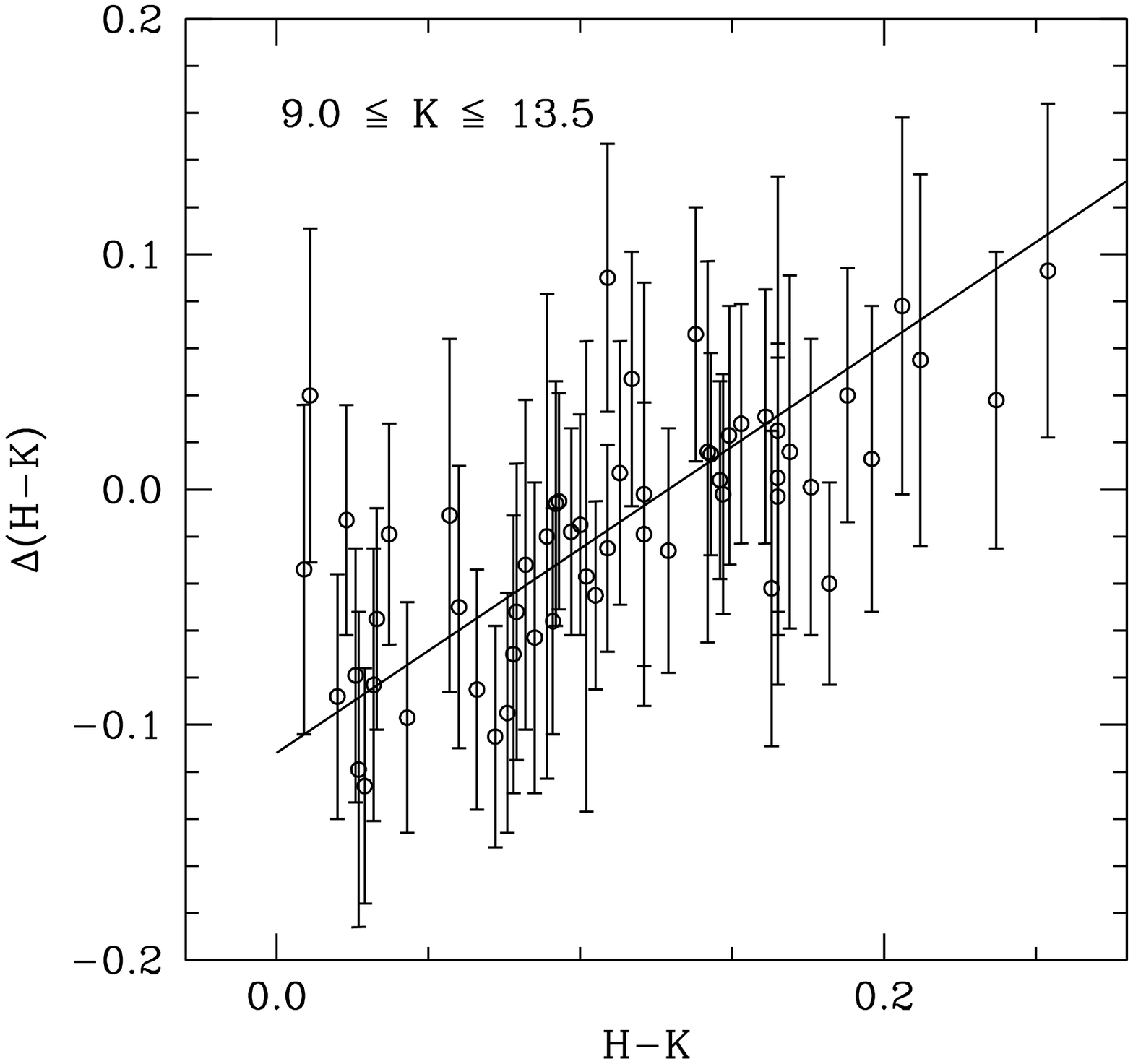]{The differences between our $H-K$ results
minus those from the 2MASS survey, after transformation to
the CIT photometric system. Only stars with highest quality
photometry and in a limited range of $K$ magnitudes are shown.
\label{fig:fig4}}

\figcaption[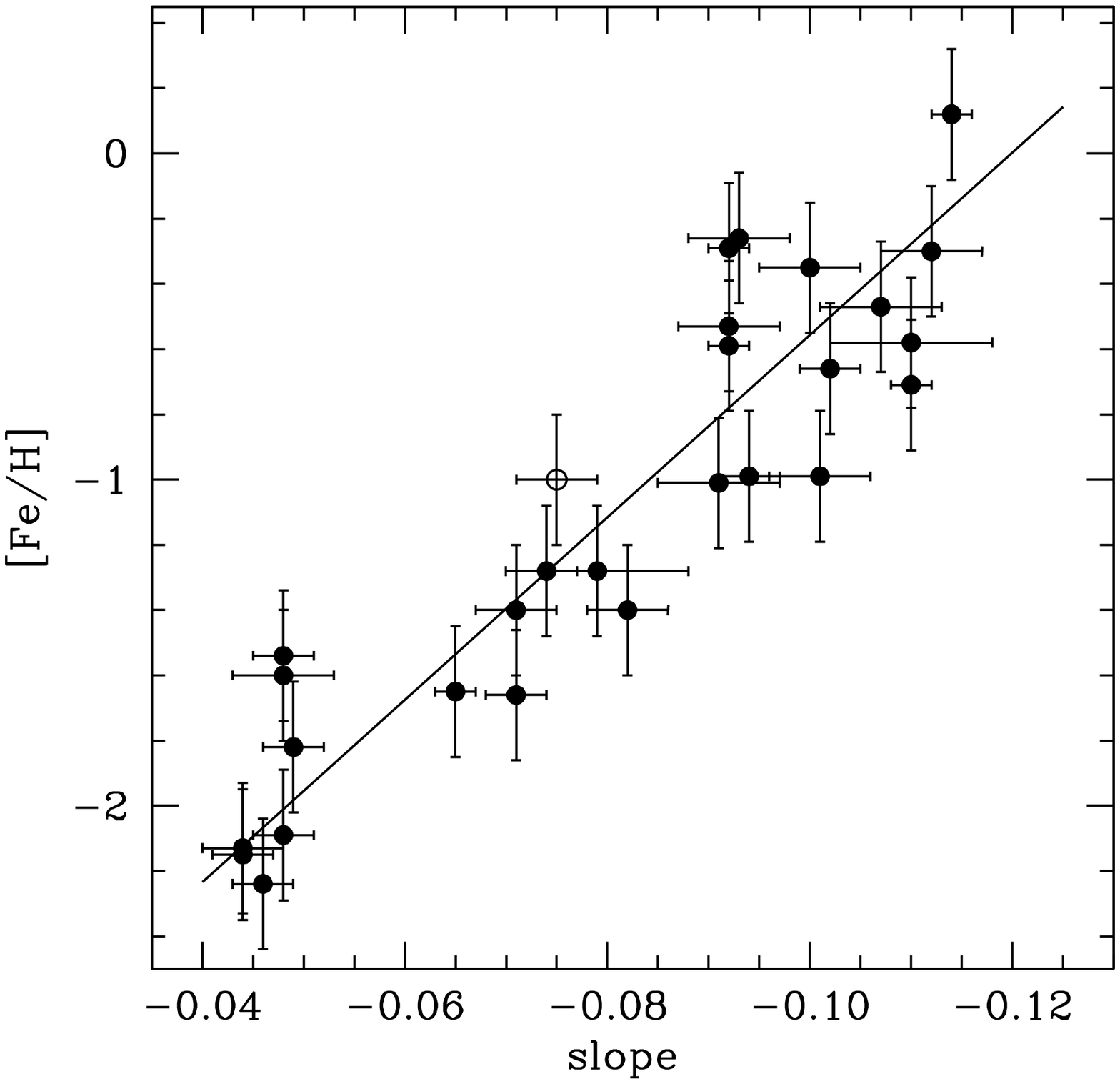]{The relation between the slope
of the red giant branch, $\Delta$($J-K$)/$\Delta K$, as a function
of [Fe/H] for globular clusters. Data for the calibrating clusters
are given in Table~\ref{tab:rgbslopesvsfe}. \label{fig:fig5}}

\figcaption[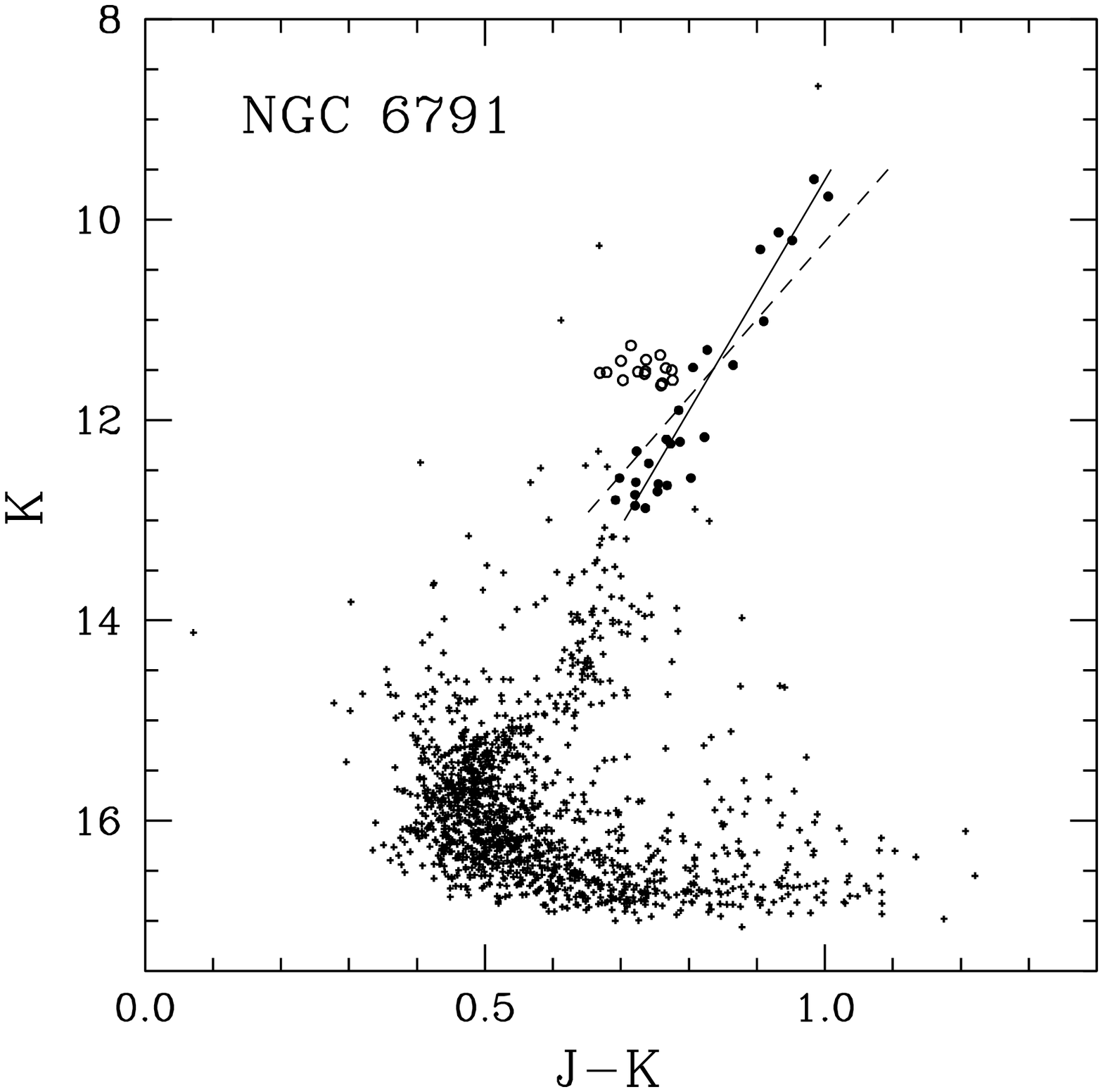]{The results of our photometry, with
red giant branch members (filled circles) and red horizontal branch
members (open circles) identified. The solid line is the ordinary
least squares bissector measurement of the slope of the red giant branch,
while the dashed line represents the expected slope for [Fe/H] = +0.3,
adopting the calibrations discussed in the text. The two lines have
been adjusted to intersect at the magnitude level of the red horizontal
branch. \label{fig:fig6}}

\figcaption[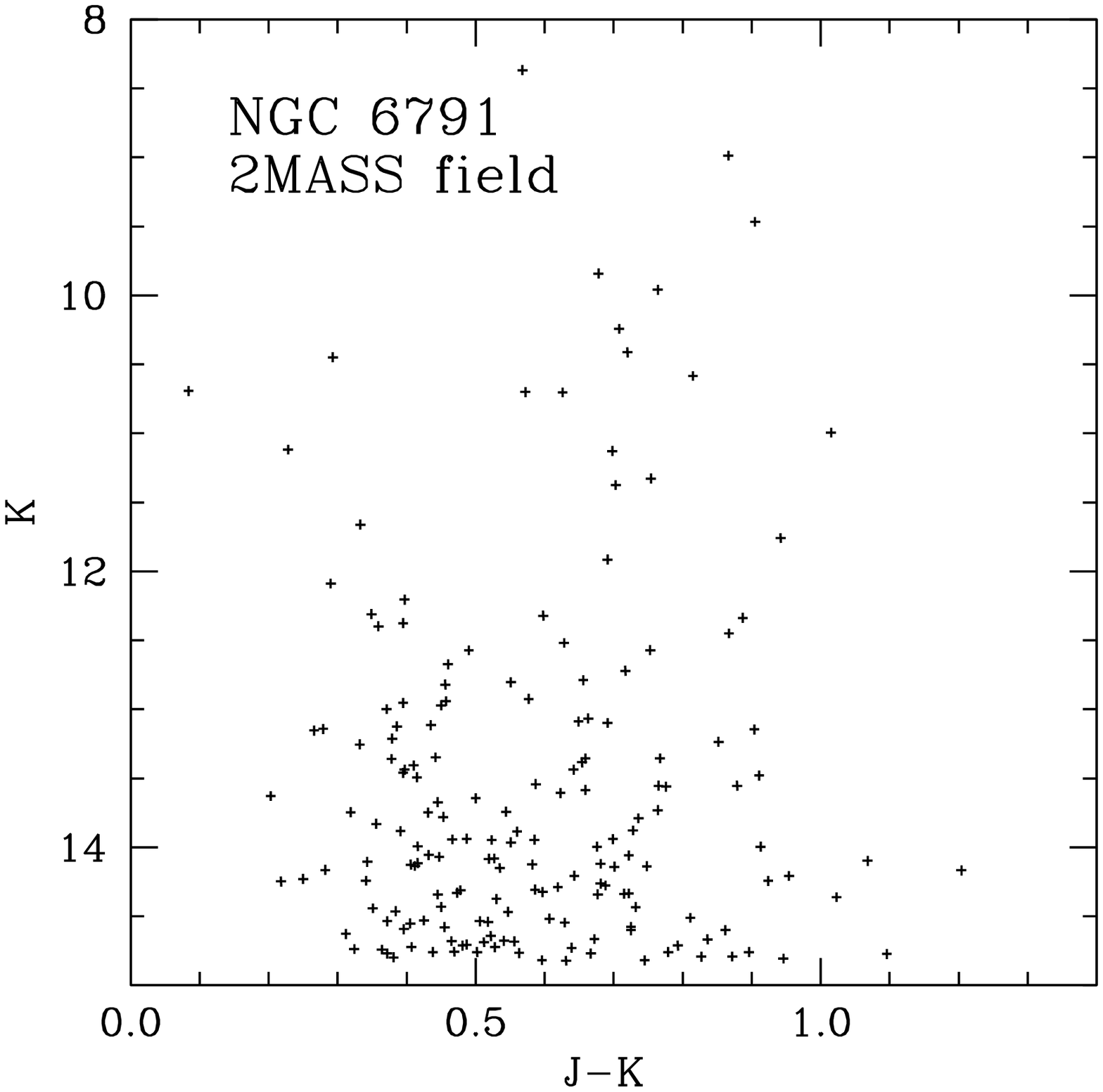]{The color-magnitude diagram for
the ``field" around NGC~6791, taken from the 2MASS survey.
Four separation regions were used to
provide the field background sample. Each region has
a radius of 3\arcmin, and each is centered 15\arcmin\ away
from the cluster center along north, south, east, and west
directions. \label{fig:fig7}}

\figcaption[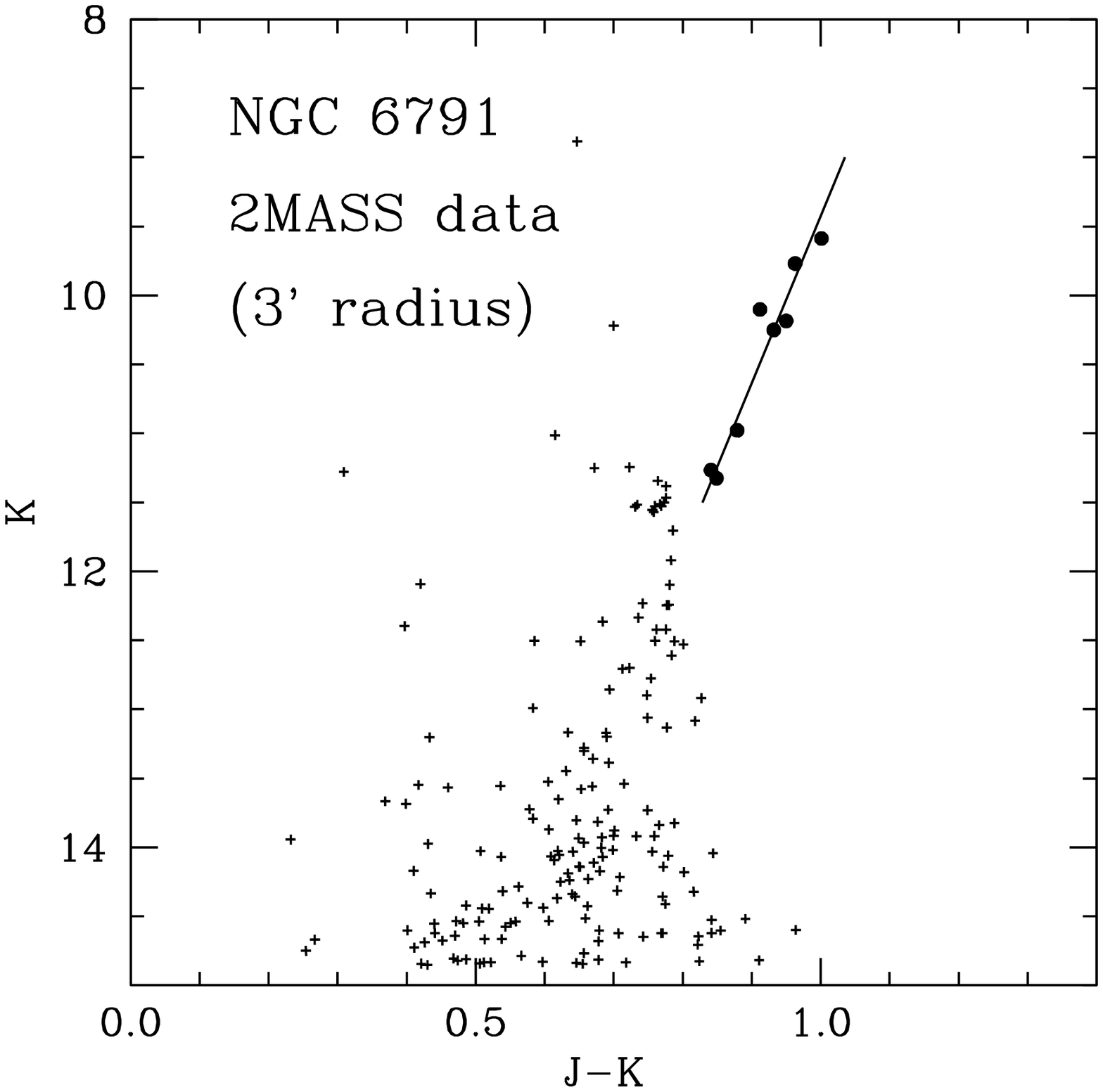]{The color-magnitude
diagram of NGC~6791 obtained using 2MASS data. In this
case, the stars were selected to lie within 3\arcmin\ of the
cluster center. The filled circles represent stars used
to measure the slope of the red giant branch, indicated by
the solid line, using an ordinary least squares bissector
analysis. \label{fig:fig8}}

\figcaption[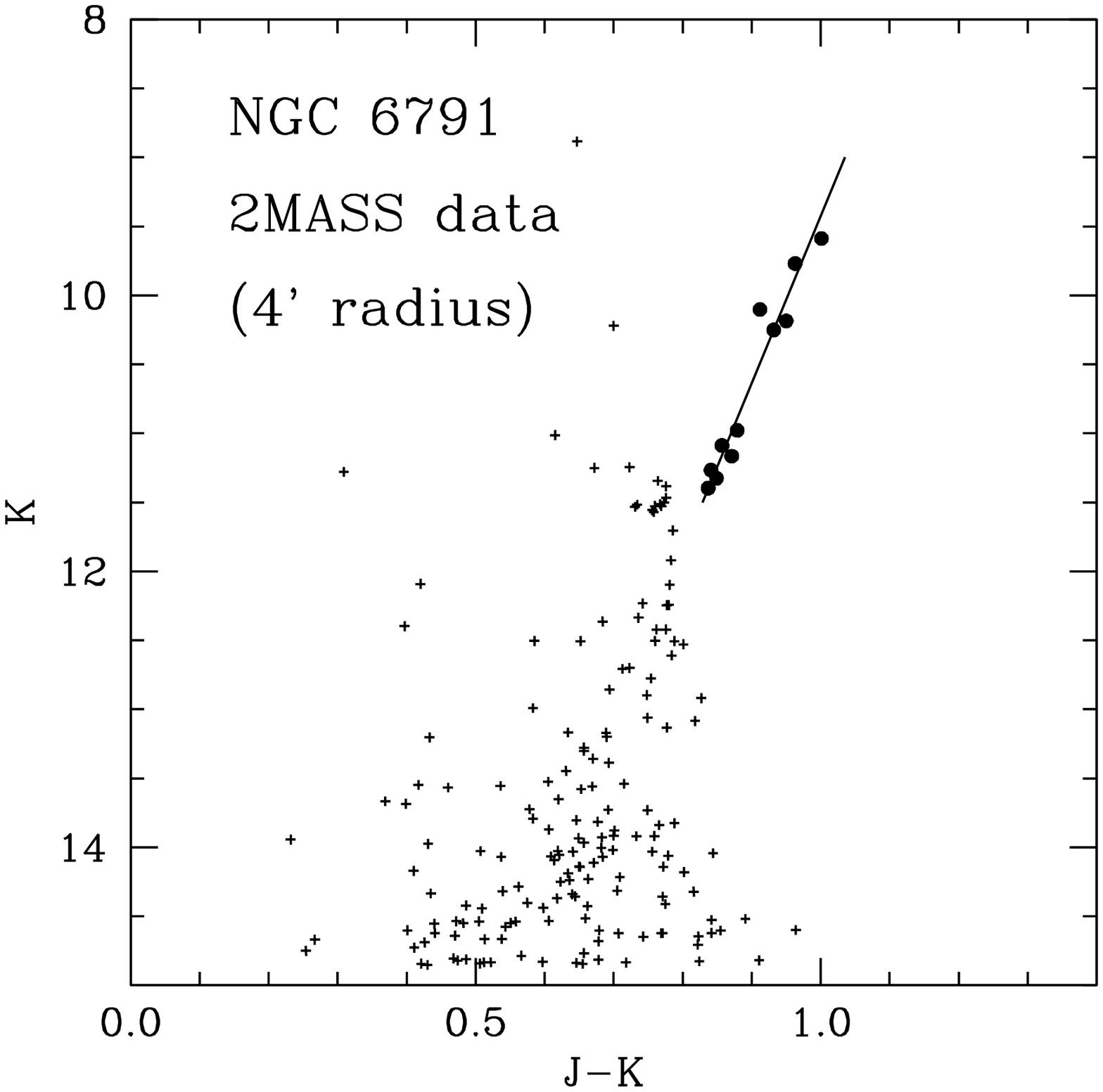]{The color-magnitude
diagram of NGC~6791 obtained using 2MASS data. In this
case, the stars were selected to lie within 4\arcmin\ of the
cluster center. The filled circles represent stars used
to measure the slope of the red giant branch, indicated by
the solid line, using an ordinary least squares bissector
analysis. \label{fig:fig9}}

\figcaption[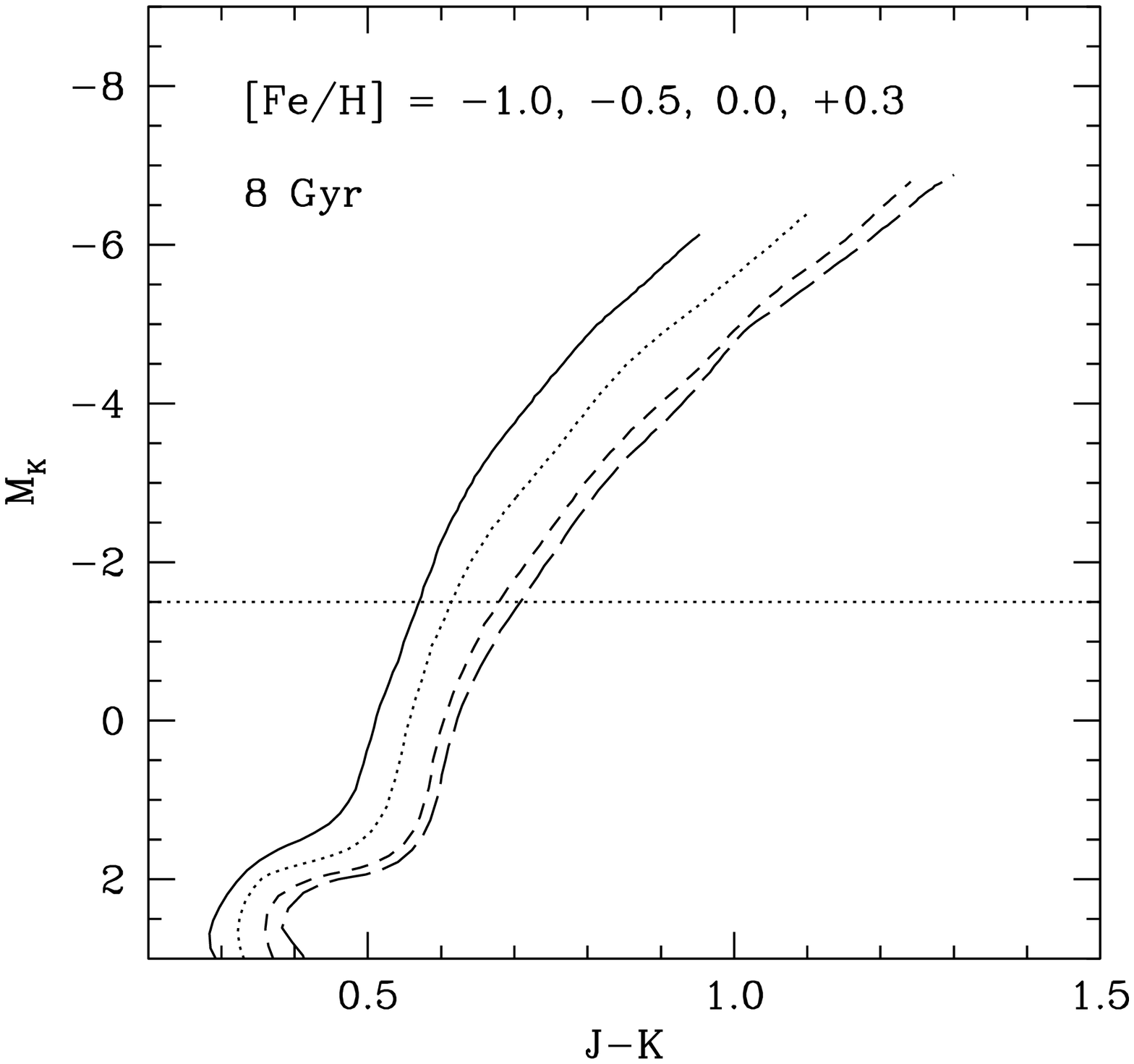]{Theoretical isochrones of $M_{K}$ vs.\
$J-K$ from YKD2003. Note that the color and slope of the
red giant branch are sensitive to [Fe/H], but that the
slope is {\em not} linear. \label{fig:fig10}}

\figcaption[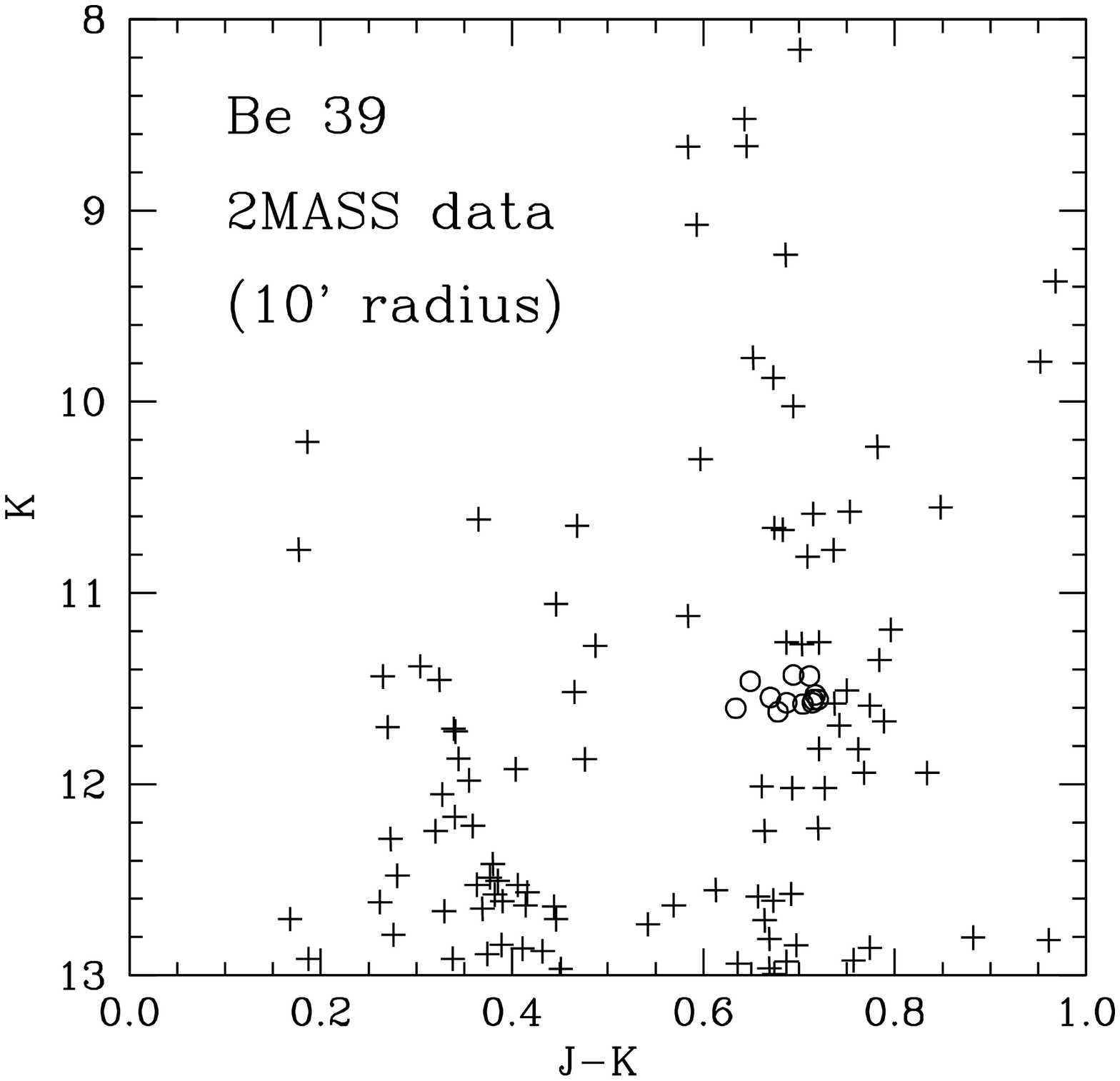]{The color-magnitude diagram of Be~39,
obtain from the 2MASS database, and restricted a region
within 10\arcmin\ of the cluster's center. The stars we
identify as red horizontal branch or red clump stars are
indicated as open circles. \label{fig:fig11}}

\figcaption[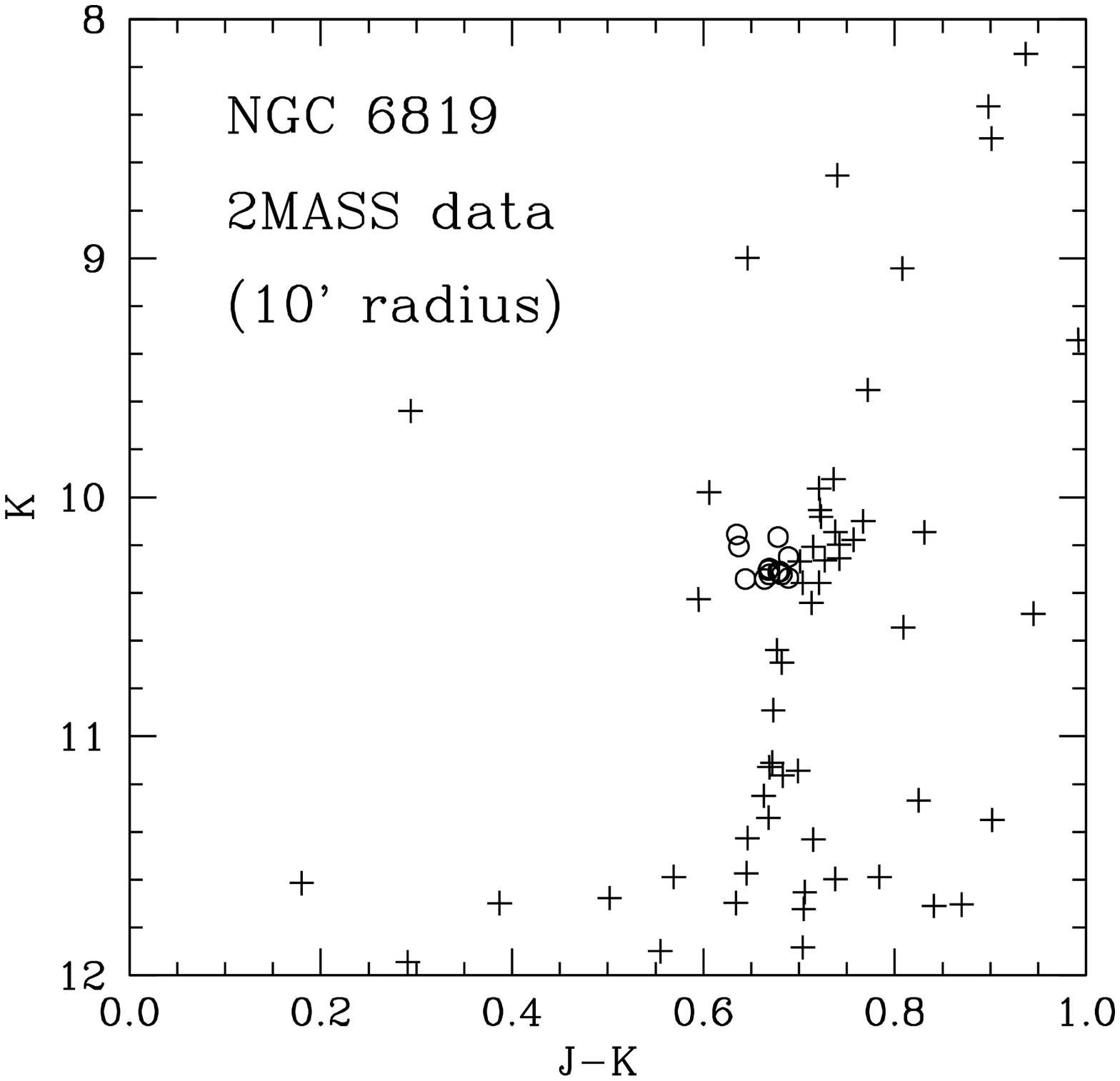]{The color-magnitude diagram of NGC~6819,
obtain from the 2MASS database, and restricted a region
within 5\arcmin\ of the cluster's center. The stars we
identify as red horizontal branch or red clump stars are
indicated as open circles. \label{fig:fig12}}

\figcaption[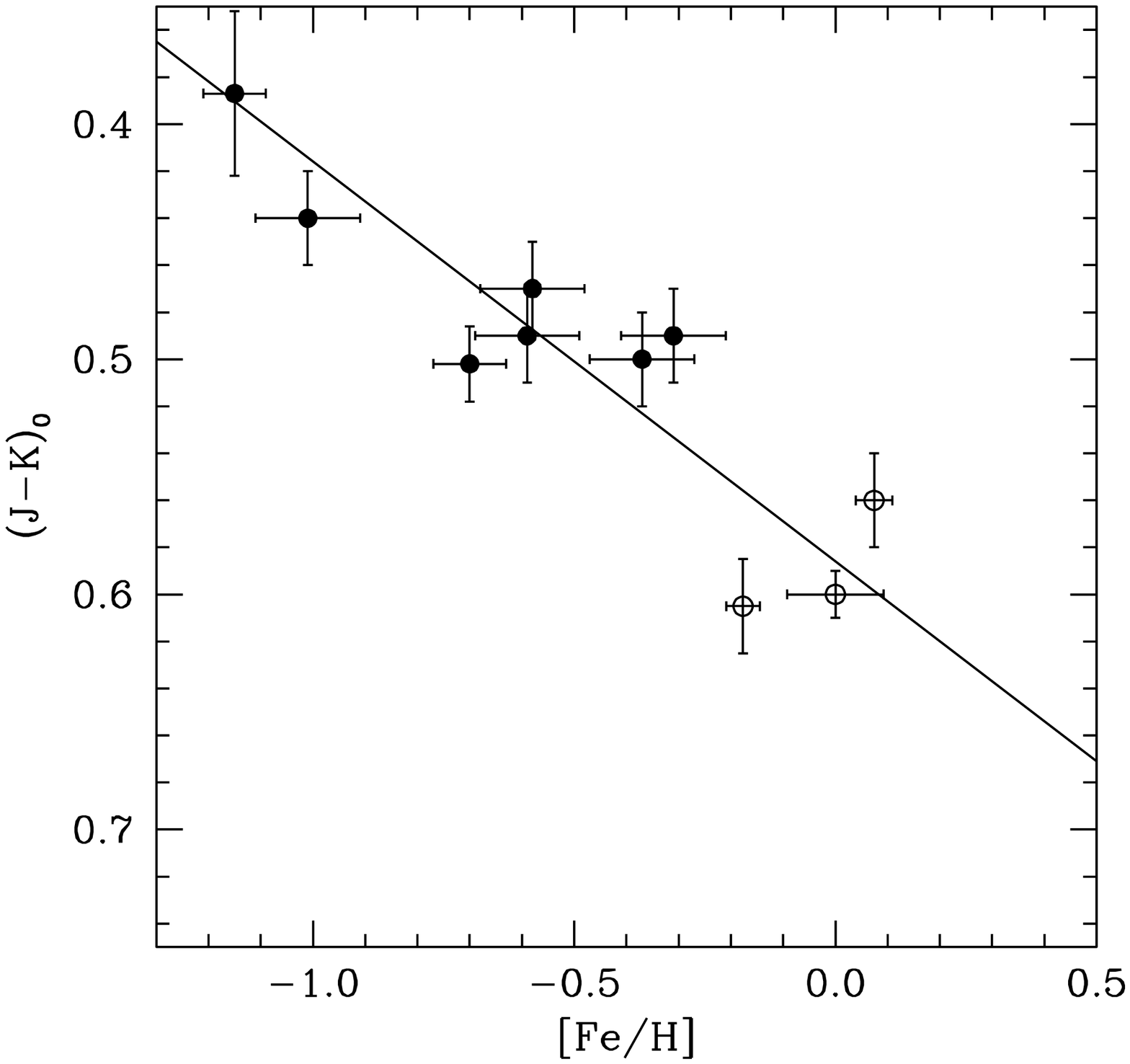]{The de-reddened mean $J-K$ colors (in the
CIT photometric system) of the red horizontal branch stars in
the oldest open clusters and globular clusters. \label{fig:fig13}}

\figcaption[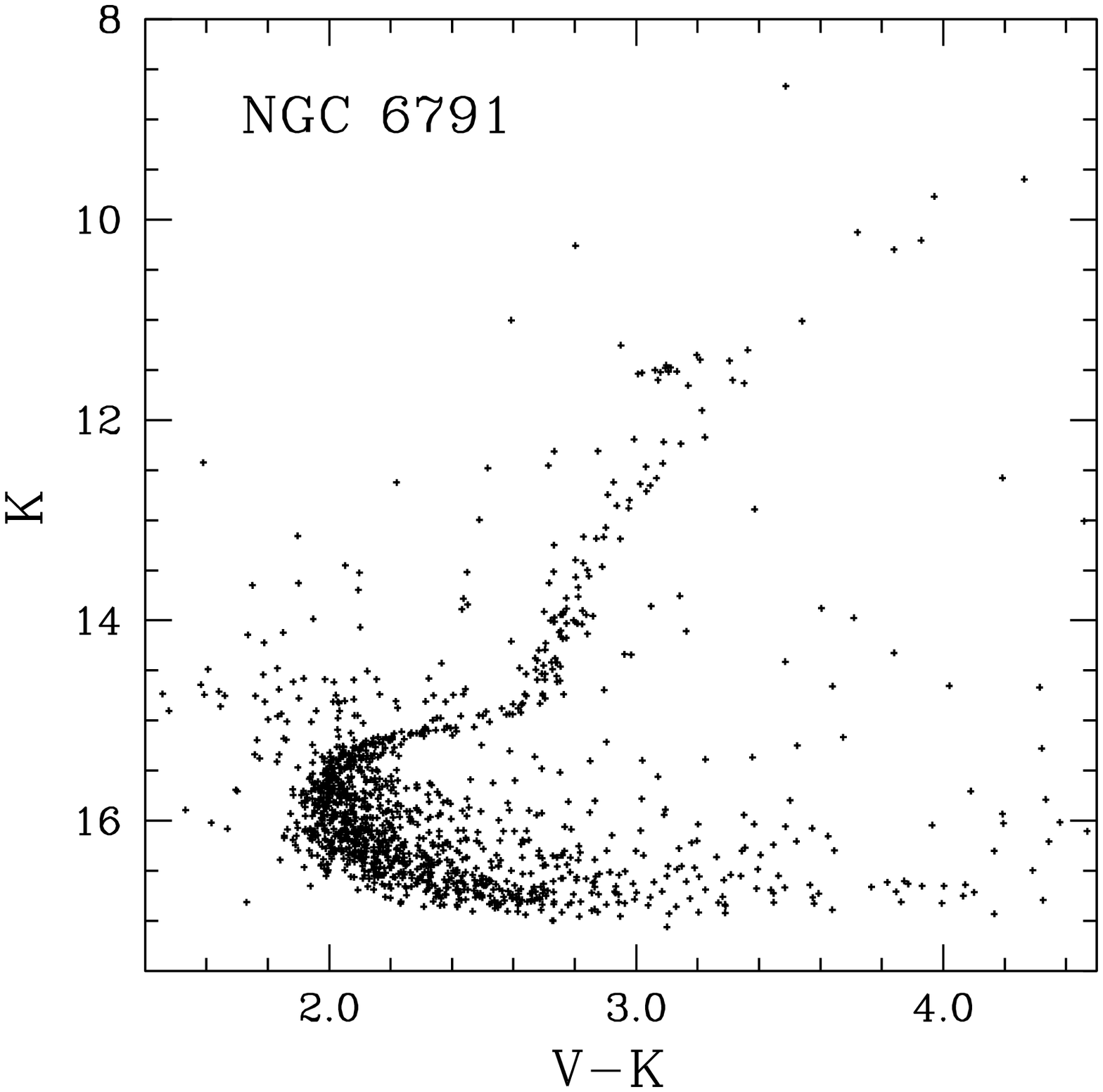]{The combined optical and infrared
photometry for stars in our observing program, after outliers in
the $J-K$ vs.\ $V-K$ diagram have been removed. Here we compare
an infrared magnitude and an optical-infrared color. Notice that
the shape of the red giant branch is distinctly curved compared
to Figure~\ref{fig:fig1}.\label{fig:fig14}}

\figcaption[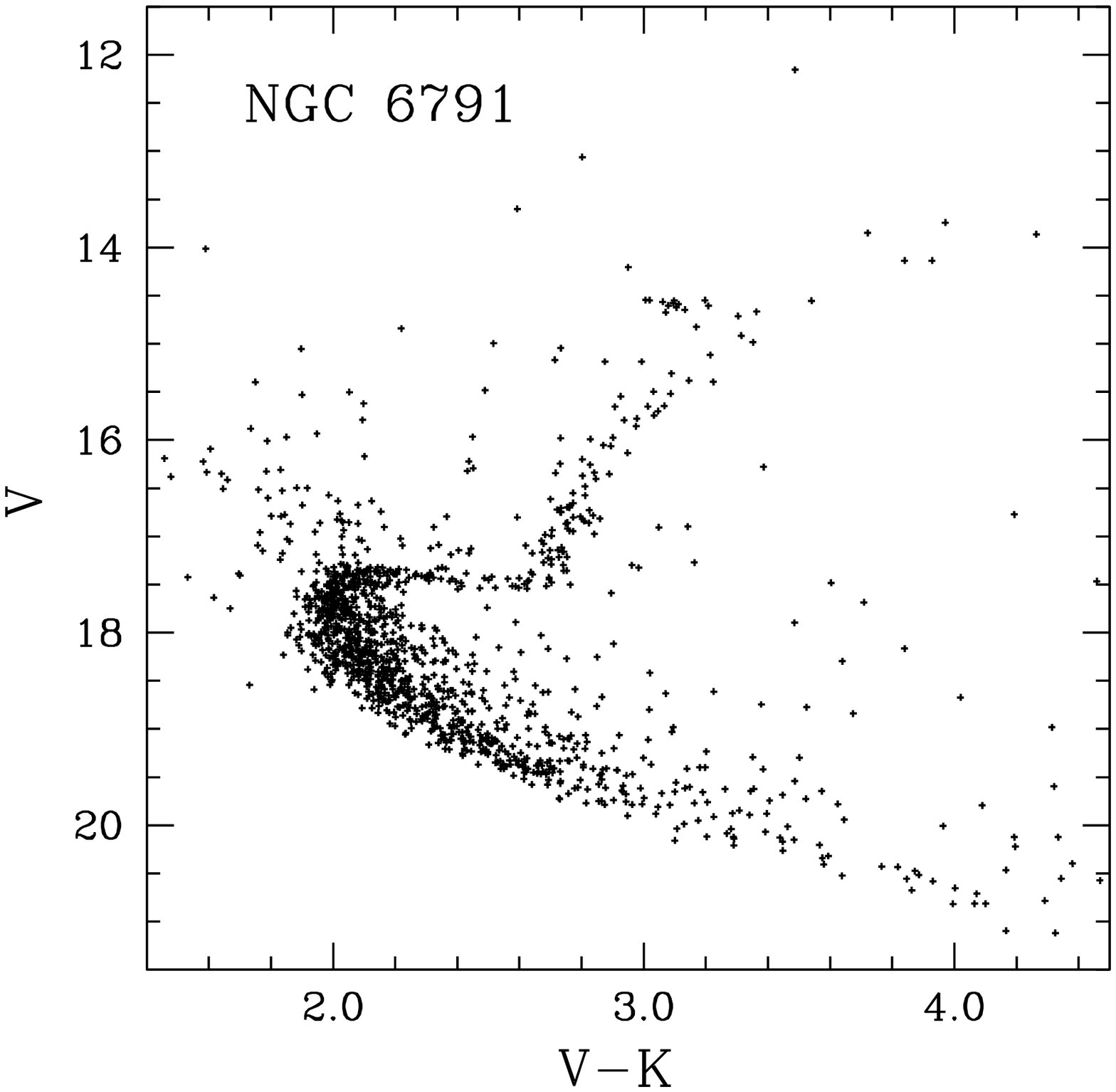]{The combined optical and infrared
photometry for stars in our observing program, after outliers in
the $J-K$ vs.\ $V-K$ diagram have been removed. Here we compare
an optical magnitude and an optical-infrared color. Notice that
the shape of the red giant branch is distinctly curved even more
than in Figure~\ref{fig:fig1} or Figure~\ref{fig:fig14}.
\label{fig:fig15}}

\figcaption[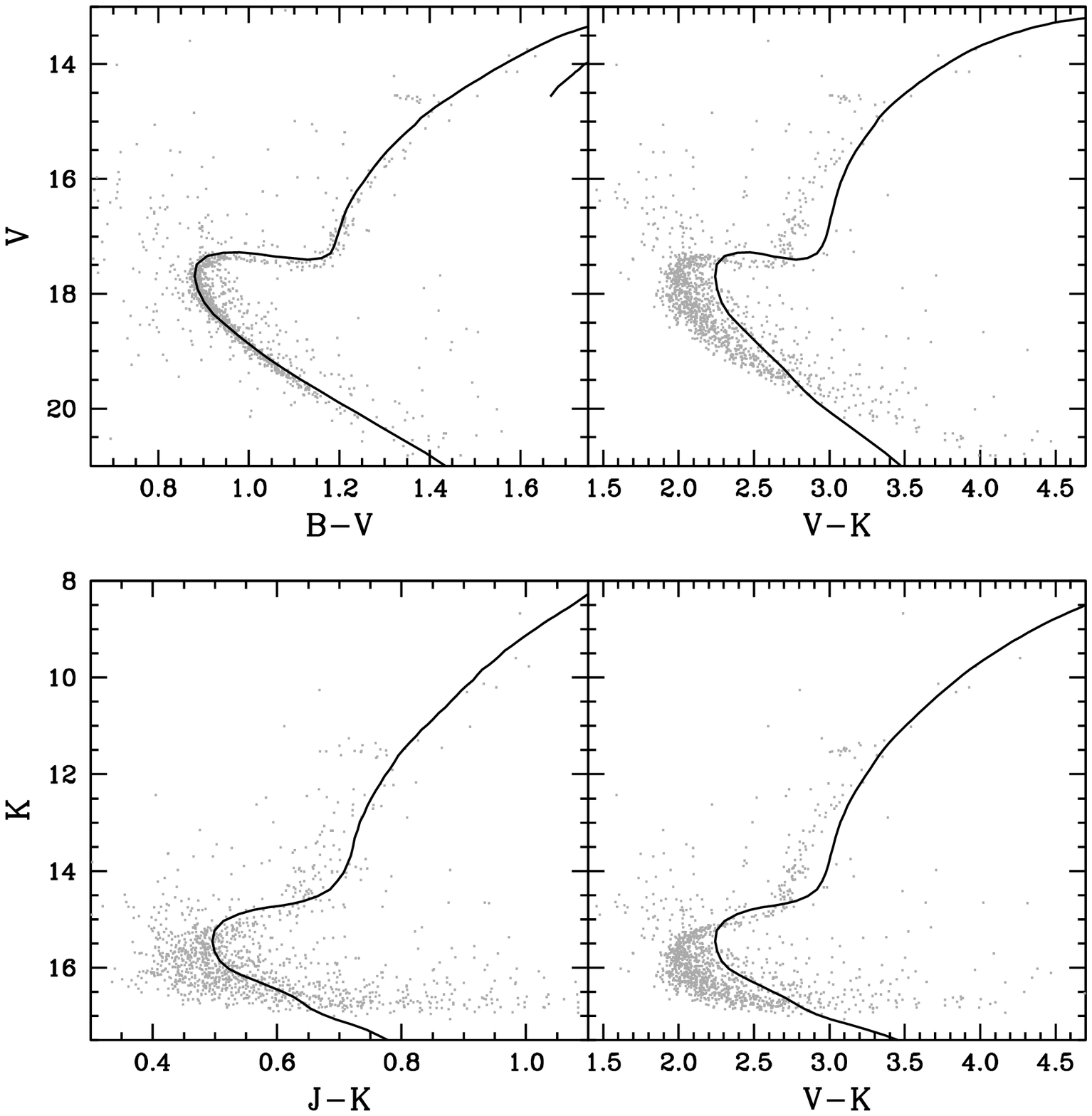]{Here we show a comparison between model
isochrones taken from YKD2003 and the optical and infrared photometry.
We first identify an optimum match in $V$ vs.\ $B-V$, using the
optical photometry of SBG2003. In this case we found a good match
for [Fe/H] = 0.0, [$\alpha$/Fe] = 0.0, E($B-V$) = 0.26 mag,
a distance modulus of 12.73 mag, and an age of 8 Gyrs. However,
the infrared color-magnitude diagrams do not agree well at all,
implying significant errors in the reddening, hence distance, or
in the metallicity. \label{fig:fig16}}

\figcaption[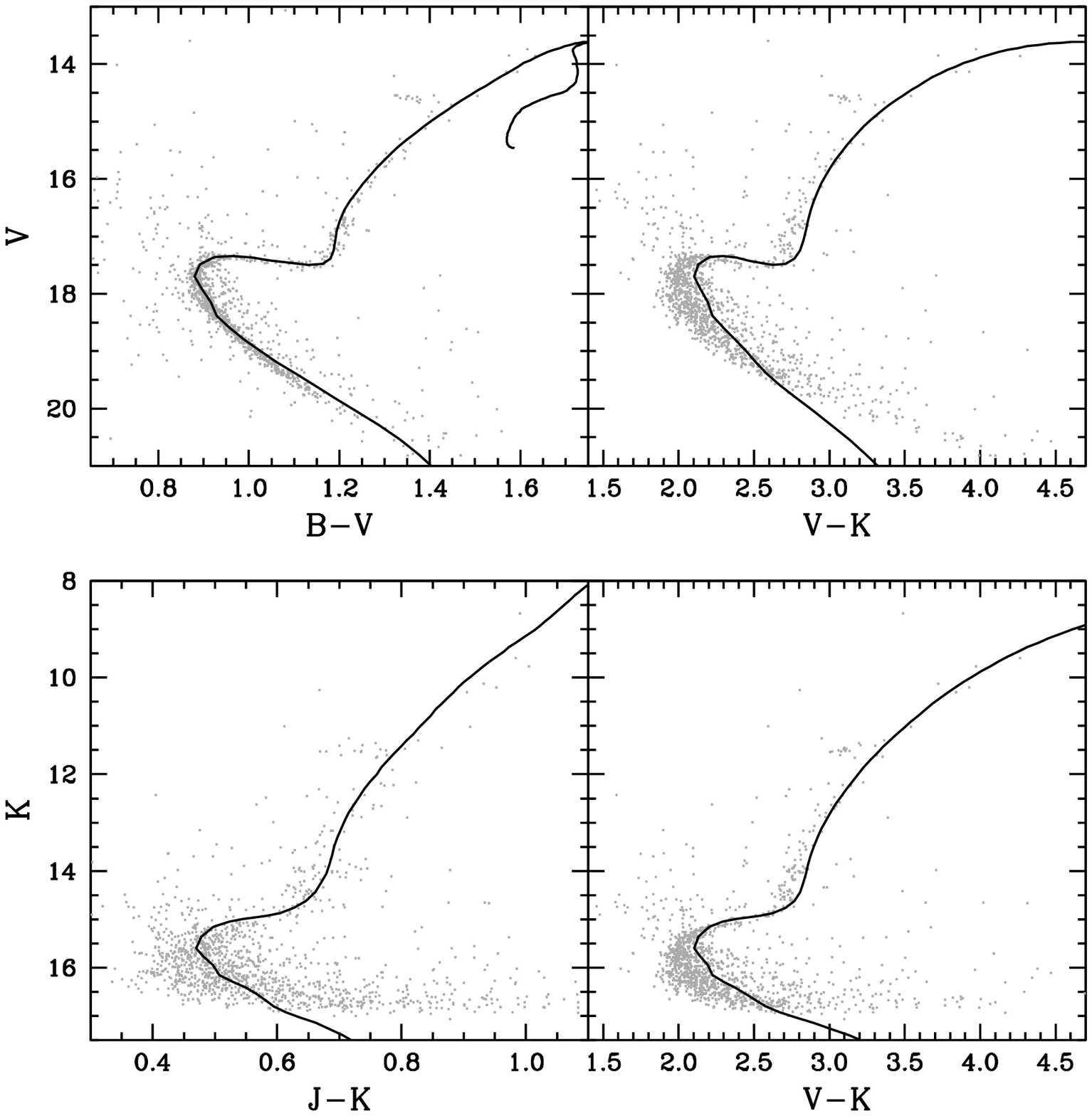]{We repeat the comparisons of
Figure~\ref{fig:fig16}, but using [Fe/H] = +0.3 and
[$\alpha$/Fe] = 0.0. Here the optimal fit to the optical data
suggests E($B-V$) = 0.17 mag, ($m-M$)$_{0}$ = 12.93 mag, and,
again, an age of 8 Gyrs. \label{fig:fig17}}

\figcaption[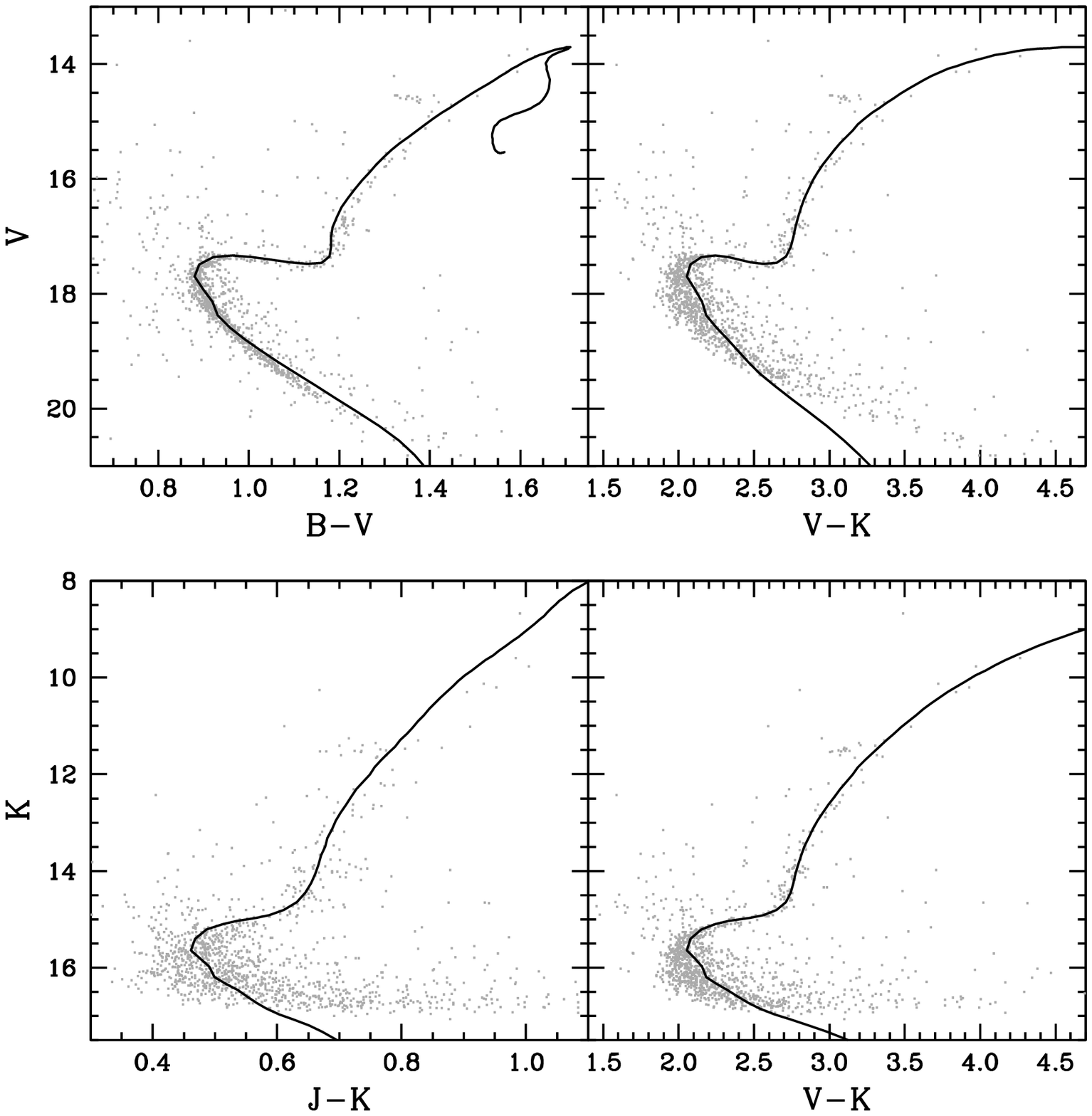]{We repeat the comparisons of
Figure~\ref{fig:fig16}, but using [Fe/H] = +0.4 and
[$\alpha$/Fe] = 0.0. Here the optimal fit to the optical data
suggests E($B-V$) = 0.13 mag, ($m-M$)$_{0}$ = 12.96 mag, and,
again, an age of 8 Gyrs. \label{fig:fig18}}

\figcaption[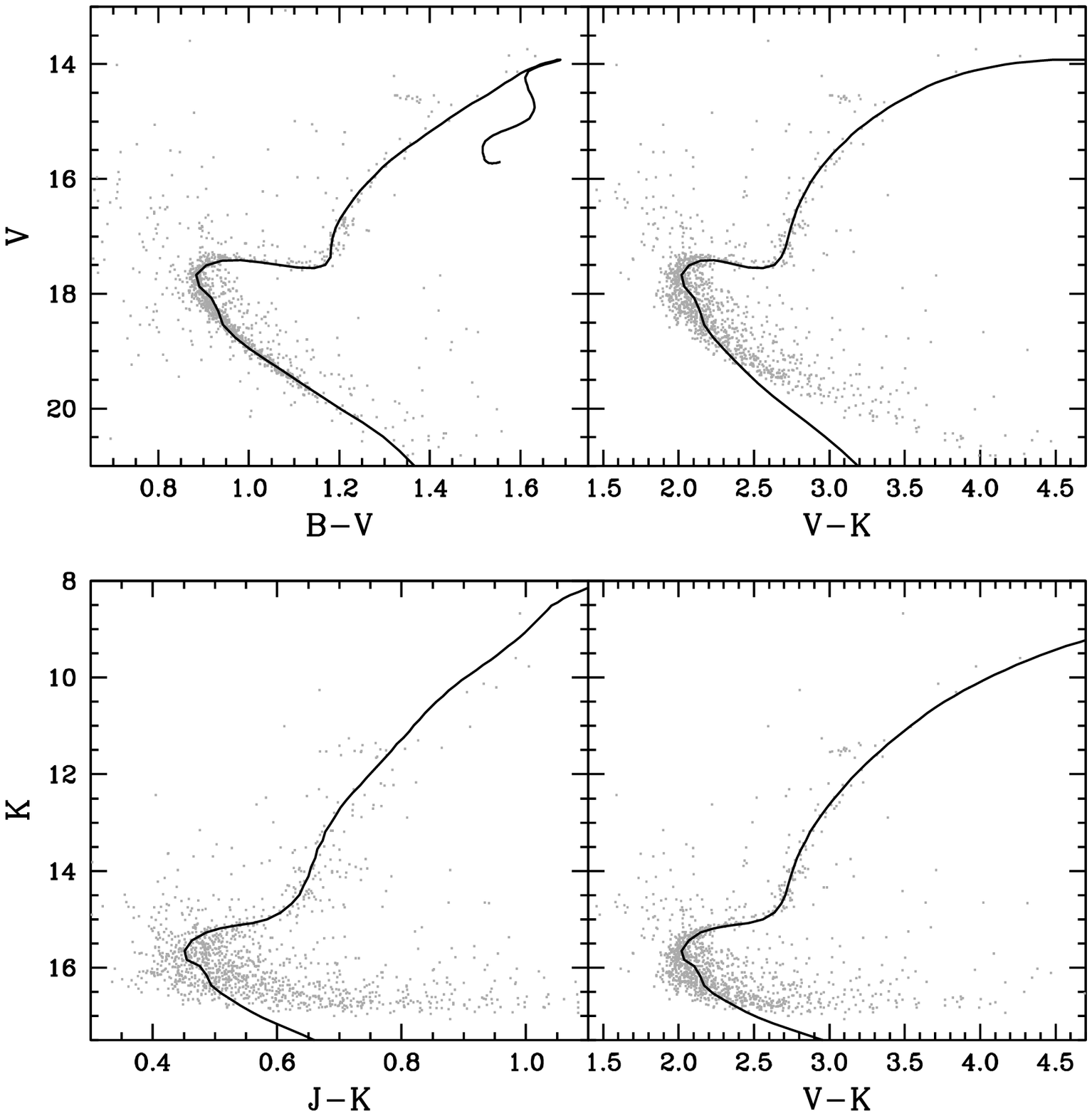]{We repeat the comparisons of
Figure~\ref{fig:fig16}, but using [Fe/H] = +0.5 and
[$\alpha$/Fe] = 0.0. Here the optimal fit to the optical data
suggests E($B-V$) = 0.11 mag, ($m-M$)$_{0}$ = 13.15 mag, and
an age of 7.5 Gyrs. \label{fig:fig19}}

\clearpage

\plotone{Lee.fig01.ps}

\clearpage

\plotone{Lee.fig02.ps}

\clearpage

\plotone{Lee.fig03.ps}

\clearpage

\plotone{Lee.fig04.ps}

\clearpage

\plotone{Lee.fig05.ps}

\clearpage

\plotone{Lee.fig06.ps}

\clearpage

\plotone{Lee.fig07.ps}

\clearpage

\plotone{Lee.fig08.ps}

\clearpage

\plotone{Lee.fig09.ps}

\clearpage

\plotone{Lee.fig10.ps}

\clearpage

\plotone{Lee.fig11.ps}

\clearpage

\plotone{Lee.fig12.ps}

\clearpage

\plotone{Lee.fig13.ps}

\clearpage

\plotone{Lee.fig14.ps}

\clearpage

\plotone{Lee.fig15.ps}

\clearpage

\plotone{Lee.fig16.ps}

\clearpage

\plotone{Lee.fig17.ps}

\clearpage

\plotone{Lee.fig18.ps}

\clearpage

\plotone{Lee.fig19.ps}

\end{document}